\documentclass[aps,amsmath,notitlepage,twocolumn,amssymb,prx,longbibliography]{revtex4-1}

\usepackage{graphicx}
\usepackage{dcolumn}
\usepackage{bm}
\usepackage[colorlinks=true,linkcolor = blue,citecolor=blue,urlcolor=blue,anchorcolor = blue]{hyperref}
\usepackage{wasysym}
\usepackage{stmaryrd}
\usepackage{verbatim}
\usepackage{subfigure}
\usepackage{amsmath}
\newcommand{\RNum}[1]{\uppercase\expandafter{\romannumeral #1\relax}}

\begin{document}

\title{Topological thermal Hall effect of ``magnetic monopoles''\\ 
in pyrochlore U(1) spin liquid}
\author{Xiao-Tian Zhang$^{1,2}$}
\author{Yong Hao Gao$^{3,1}$}
\author{Chunxiao Liu$^{4}$}
\author{Gang Chen$^{1,3,5}$}
\email{gangchen.physics@gmail.com}
\affiliation{$^{1}$Department of Physics and HKU-UCAS Joint 
Institute for Theoretical and Computational Physics at Hong Kong, 
The University of Hong Kong, Hong Kong, China}
\affiliation{$^{2}$International Center for Quantum Materials, 
Peking University, Beijing, 100871, China}
\affiliation{$^{3}$State Key Laboratory of Surface Physics and Department of Physics, 
Fudan University, Shanghai 200433, China}
\affiliation{$^{4}$Department of Physics, University of California, 
Santa Barbara, California 93106, USA}
\affiliation{$^{5}$Collaborative Innovation Center of Advanced Microstructures, 
Nanjing University, Nanjing 210093, China}

\date{\today}

\begin{abstract}
``Magnetic monopole'' is an exotic quantum excitation in three dimensional U(1) spin liquid, 
and its emergence is purely of quantum origin and has no classical analogue. 
We predict topological thermal Hall effect (TTHE) of ``magnetic monopoles''
and present this prediction through non-Kramers doublets on a pyrochlore lattice. 
We observe that, when the external magnetic field polarizes the Ising component of the local 
moment, internally this corresponds to the induction of emergent dual U(1) gauge 
flux for the ``magnetic monopoles''. The motion of ``magnetic monopoles'' is 
then twisted by the induced dual U(1) gauge flux. This emergent Lorentz force on  
``magnetic monopoles'' is the fundamental origin of TTHE. Therefore, TTHE would 
be a direct evidence of the ``monopole''-gauge coupling and the emergent U(1) gauge 
structure in pyrochlore U(1) spin liquid. Our result does not depend strongly on our
choice of non-Kramers doublets for our presentation, and can be well extended to 
Kramers doublets. Our prediction can be readily tested among the pyrochlore spin 
liquid candidate materials. We give a detailed discussion about the expectation 
for different pyrochlore magnets. 
\end{abstract}

\maketitle

\section{Introduction}
\label{sec1}

Emergent gauge structure and theory comprise an important 
subject in modern condensed matter physics, particularly 
for strongly correlated quantum matter~\cite{XGWenbook}. 
It is this theory that underlies the unified gauge theory 
description of fractional quantum Hall effect and quantum 
spin liquids (QSLs)~\cite{XGWenbook}. 
While an initial understanding of the fractional quantum Hall effect(FQHE) relies 
on Laughlin's construction of a variational wave-function~\cite{Laughlin}, later on, 
Ginzburg-Landau field theoretical descriptions are developed 
conceiving an additional gauge interaction described by the Chern-Simons gauge 
theories~\cite{PhysRevLett.62.82,PhysRevB.44.5246}.
The discovery of QSLs follows a completely independent line 
of development pioneered by Anderson and collaborators~\cite{Anderson1973,ANDERSON1987,Baskaran1987}.
Intriguingly, a QSL state, dubbed as `chiral spin liquid' state, 
is proposed to be equivalent to the FQHE~\cite{PhysRevLett.59.2095}.
And, the modern understanding of QSLs has been greatly advanced by 
various lattice gauge theories~\cite{BalentsS,PhysRevB.62.7850,PhysRevB.69.064404} 
conceiving non-local, fractionalized excitations. 
To confirm the existence of QSLs in a realistic quantum material, 
one has to establish the presence of the emergent gauge structure and 
the associated fractionalized quantum particles, 
e.g., the spinon and ``magnetic monopole'' in U(1) QSL.
This requires a mutual feedback between theories and experiments. 
More precisely, one needs to understand how the emergent gauge structure 
manifests itself in the actual experimental observables. 
In a more progressive manner, it would be beneficial to provide 
some level of controllability or prediction 
of these emergent phenomena from the understanding of the relationship 
between the microscopic physics and the emergent gauge structure. 
In this effort, some of us have proposed ways to 
spectroscopically control the spinon band structure 
and then the spinon continuum in the inelastic neutron scattering measurement 
for several QSL candidates~\cite{PhysRevB.95.041106,PhysRevB.96.075105,PhysRevB.96.054445,Chen1902.07075} 
such as Ce$_2$Sn$_2$O$_7$, Ce$_2$Zr$_2$O$_7$ 
and YbMgGaO$_4$~\cite{sibille1502candidate,dai1901experimental,YueshengSR,
YShenNcomm,YMGOYao,PhysRevLett.117.097201,PhysRevX.8.031001,nphys2017,YueshengPRL1}.
As for the transport properties, 
 two of us have further studied the strong Mott insulating QSLs and suggested the origin of the emergent 
Lorentz force from the antisymmetric Dzyaloshinskii-Moriya interaction 
for the spinons as the source of the topological thermal Hall 
conductivity in these systems~\cite{Chen1901.01522,PhysRevResearch.1.013014}. 
In this paper, we turn our attention to 
study the thermal Hall transport in another important QSL state, 
namely the pyrochlore U(1) QSL.

\begin{figure}[b]
\centering
\includegraphics[width=7cm]{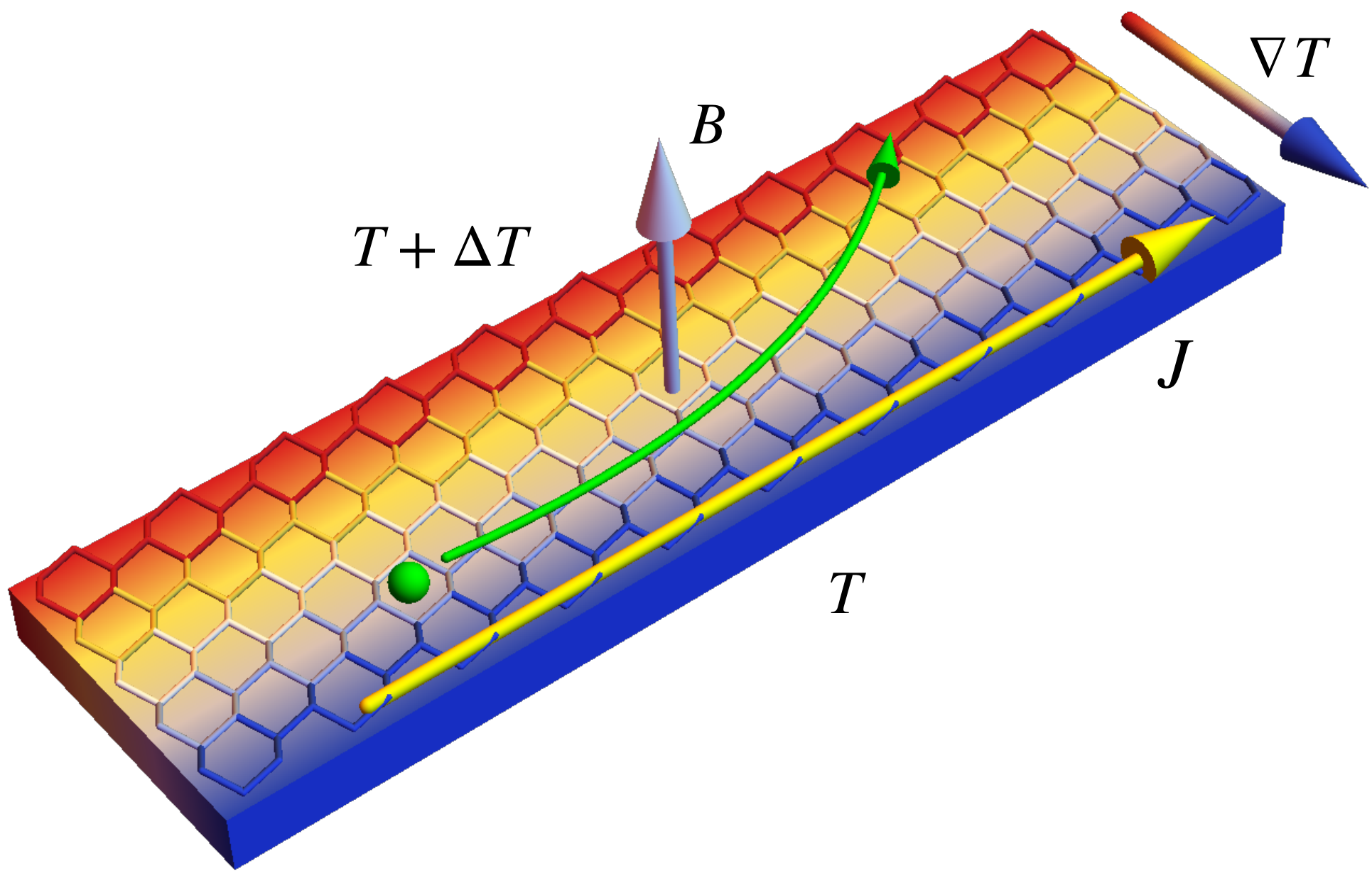}
\caption{(Color online.) Schematic picture of the thermal Hall effect from the 
``magnetic monopoles'' on the dual diamond lattice for the pyrochlore U(1) QSL, 
where the heat current ``$J$'' has contributions from all mobile excitations. 
We single out the ``magnetic monopoles'' (in green) that are suggested to contribute to 
the thermal Hall effect in this work. } 
\label{1}
\end{figure}

The pyrochlore U(1) QSL is described by the emergent compact U(1) lattice gauge theory,
and supports the gapless U(1) gauge photon, gapped spinon and ``magnetic monopole'' as 
its elementary excitations~\cite{PhysRevB.69.064404}. 
Many pyrochlore materials, 
mainly the rare-earth pyrochlores~\cite{Gingras2014,PhysRevB.86.104412,PhysRevLett.98.157204,PhysRevB.83.094411,PhysRevB.86.075154,PhysRevX.7.041057,BalentsS}, 
have been proposed as candidates to realize this U(1) QSL~\citep{BalentsS}. 
Although many interesting experimental signatures have been suggested, 
the firm establishment of pyrochlore U(1) QSL has not yet been settled
for any material. In this paper, we develop a theory to predict the phenomenon  
of the topological thermal Hall effect (TTHE) in the pyrochlore U(1) QSL and 
propose it as a positive evidence of the emergent U(1) gauge structure. 
Our observation stems from the physical meaning of the spin variables 
in the U(1) QSL. It is observed that, the Ising component of the spin
works as an emergent electric field in the U(1) lattice gauge theory. 
From the view of the dual gauge theory, this emergent and internal electric 
field behaves as a dual U(1) gauge flux for the ``magnetic monopoles.'' 
The {\sl external magnetic field}, that couples linearly with the spins
through a simple Zeeman coupling, polarizes the internal electric field 
and thereby modifies the dual U(1) gauge flux that is experienced by the 
``magnetic monopoles.'' This coupling between the internal variable and 
the external field effectively generates an emergent Lorentz force on 
the ``magnetic monopoles'' and creates a TTHE in the system. The dual 
Hamiltonian for the ``magnetic monopoles,'' that captures this effect, 
is given as  
\begin{equation}
\begin{aligned}
{\cal H}_{\rm dual}=
& - t\sum_{ \langle \textsf{r} \textsf{r}^\prime \rangle } 
\Phi^\dagger_{\textsf{r}} \Phi_{\textsf{r}^\prime} 
e^{-i 2\pi a_{\textsf{r}\textsf{r}^\prime}}
-\mu \sum_{\textsf{r}} 
\Phi^\dagger_{\textsf{r}} \Phi^{}_{\textsf{r}}
\\
& + \sum_{rr^\prime} \frac{U}{2} 
({\rm curl}a -\bar{E}_{rr^\prime})^2 
- K \sum_{ \textsf{r} \textsf{r}^\prime}  
\cos B_{\textsf{r} \textsf{r}^\prime} ,
\end{aligned}
\label{dual_ham}
\end{equation}
where $\Phi^\dagger_{\textsf{r}}$ $(\Phi_{\textsf{r}} )$ denotes a creation (annihilation) operator 
of the ``magnetic monopoles'' on a dual diamond lattice $\textsf{r}$-site. Here the 
sherif symbol $\textsf{r}$ is reserved for the dual diamond lattice that will be explain later. 
The first line describes the hopping of the ``magnetic monopoles'' on
the dual diamond lattice and minimally couples to the dual dynamical U(1)
gauge field $a_{\textsf{r}\textsf{r}^\prime}$, and the second line is the Maxwell term of the U(1) gauge field.
The detailed description of the notation in Eq.~\eqref{dual_ham} is given in Sec.~\ref{sec2}.
The external magnetic field modifies the dual U(1) gauge flux in the above
equation and generates the TTHE for the ``magnetic monopoles'',
which is explained in Sec.~\ref{sec3}. 

Thermal Hall effect has been measured and detected in the pyrochlore ice 
materials Tb$_2$Ti$_2$O$_7$~\cite{Hirschberger106} and Yb$_2$Ti$_2$O$_7$~\cite{Hirschberger1903.00595}. 
In Tb$_2$Ti$_2$O$_7$, the crystal electric field ground state of the Tb$^{3+}$ ion 
under the $D_{3d}$ crystal electric field is a non-Kramers doublet~\cite{PhysRevLett.98.157204}, 
although the crystal field gap to the first excited doublet 
is relatively small among the rare-earth pyrochlore magnets. 
In Yb$_2$Ti$_2$O$_7$, the crystal electric field ground state of the Yb$^{3+}$ ion is a Kramers doublet. 
In this paper, we will first deliver our theory with the non-Kramers doublets
for the pyrochlore ice U(1) QSL and then explain the extension to the 
Kramers doublets. Although we start with the spin ice manifold, our results
do not rely on the proximity of the spin ice configuration. As long 
as the pyrochlore U(1) QSL is realized, our results would be applicable, 
regardless whether the system is close or not close to the spin ice manifold.

The remaining parts of the paper are organized as follows. 
In Sec.~\ref{sec2}, we construct the dual lattice gauge theory for the pyrochlore 
U(1) QSL and introduce the ``magnetic monopole'' degrees of freedom into the formulation. 
In Sec.~\ref{sec3}, we present the induction of dual U(1) gauge flux 
through the Zeeman coupling. The thermal Hall current for the ``magnetic monopoles'' 
under a temperature gradient is analyzed in Sec.~\ref{sec4a} . In Sec.~\ref{sec4b}, 
we calculate the ``monnopole'' band dispersion from the mean-field monopole Hamiltonian 
with an induced dual U(1) gauge flux. In Sec.~\ref{sec4c} the temperature dependence 
of the thermal Hall conductivity is calculated.
We compare our results with other QSLs in Sec.~\ref{sec7} and give a detailed 
discussion about the expectation for different pyrochlore magnets. 
The details of calculation and derivation are presented in Appendices.

\begin{table}\renewcommand{\arraystretch}{1.2}
\begin{tabular}{ll}
\hline\hline
Excitations (notation 1) & Excitations (notation 2) 
\\
Spinon                   & Magnetic monopole 
\\
``Magnetic monopole''    &  Electric monopole 
\\
Gauge photon             & Gauge photon 
\\
\hline\hline	
\end{tabular}
\caption{Correspondence between two different notations for the elementary excitations
in pyrochlore U(1) QSL. ``Magnetic monopole'' is sometimes referred as visons
in some literature. Usually ``vison'' refers to the $\mathbb{Z}_2$ 
flux~\cite{PhysRevB.62.7850,PhysRevB.63.134521,PhysRevB.64.214511} 
for the $\mathbb{Z}_2$ topological order in 2+1D and is also known 
as ``m'' particle in Kitaev's toric code model~\cite{KITAEV20032}. }
\label{notation}
\end{table}

\section{``Magnetic monopoles' from dual lattice gauge theory}
\label{sec2}

There are two realistic spin models proposed for 
the pyrochlore U(1) QSL~\cite{PhysRevLett.112.167203,PhysRevX.1.021002,PhysRevB.83.094411}. 
Due to the spin-orbit entangled nature of 
the relevant rare-earth ion, the spin models are highly anisotropic. 
One of the spin models applies for usual Kramers doublets as well as non-Kramers doublets.
For instance, the ground state of the Yb$^{3+}$ ion in Yb$_2$Ti$_2$O$_7$ and Er$^{3+}$ ion in 
Er$_2$Ti$_2$O$_7$~\cite{PhysRevX.1.021002,PhysRevLett.109.167201} 
are Kramers doublets, while the ground state of the
 Pr$^{3+}$ ion in Pr$_2$Zr$_2$O$_7$~\cite{PhysRevLett.118.107206} 
and Tb$^{3+}$ ion in Tb$_2$Ti$_2$O$_7$~\cite{PhysRevB.83.094411} are non-Kramers doublets.
The other model, known as the XYZ model~\cite{PhysRevLett.112.167203,PhysRevB.95.041106}, 
applies for dipole-octuple doublets, such as Nd$^{3+}$ ion in 
Nd$_2$Zr$_2$O$_7$~\cite{Petit2016} and Ce$^{3+}$ ion in Ce$_2$Sn$_2$O$_7$ 
and Ce$_2$Zr$_2$O$_7$~\cite{sibille1502candidate,dai1901experimental}. 
It is known that both spin models reduce to a XXZ model in certain limit, and
 the XXZ model on a pyrochlore lattice supports a pyrochlore 
quantum ice U(1) QSL~\cite{PhysRevB.69.064404}.
Generically, this QSL state is a stable phase 
derived from the generic spin models. Although theoretical 
approaches are valid in the Ising regime~\cite{PhysRevB.69.064404}, the 
stability of the pyrochlore U(1) QSL goes beyond the perturbative 
Ising regime~\cite{PhysRevB.86.104412}. Therefore, 
we adopt a more inclusive notion of ``pyrochlore U(1) QSL''. 
In this section, we first start from the ring exchange model that is obtained from 
the realistic spin model by the degenerate perturbation theory 
in the Ising limit. The discussion is on a generic ground where the local moment 
is not specified to be a Kramers doublet or non-Kramers doublet.
Then, we obtain a lattice gauge theory and expose the ``monopoles'' explicitly
by means of electromagnetic duality transformation.

\begin{figure}[t]
\centering
\includegraphics[width=6cm]{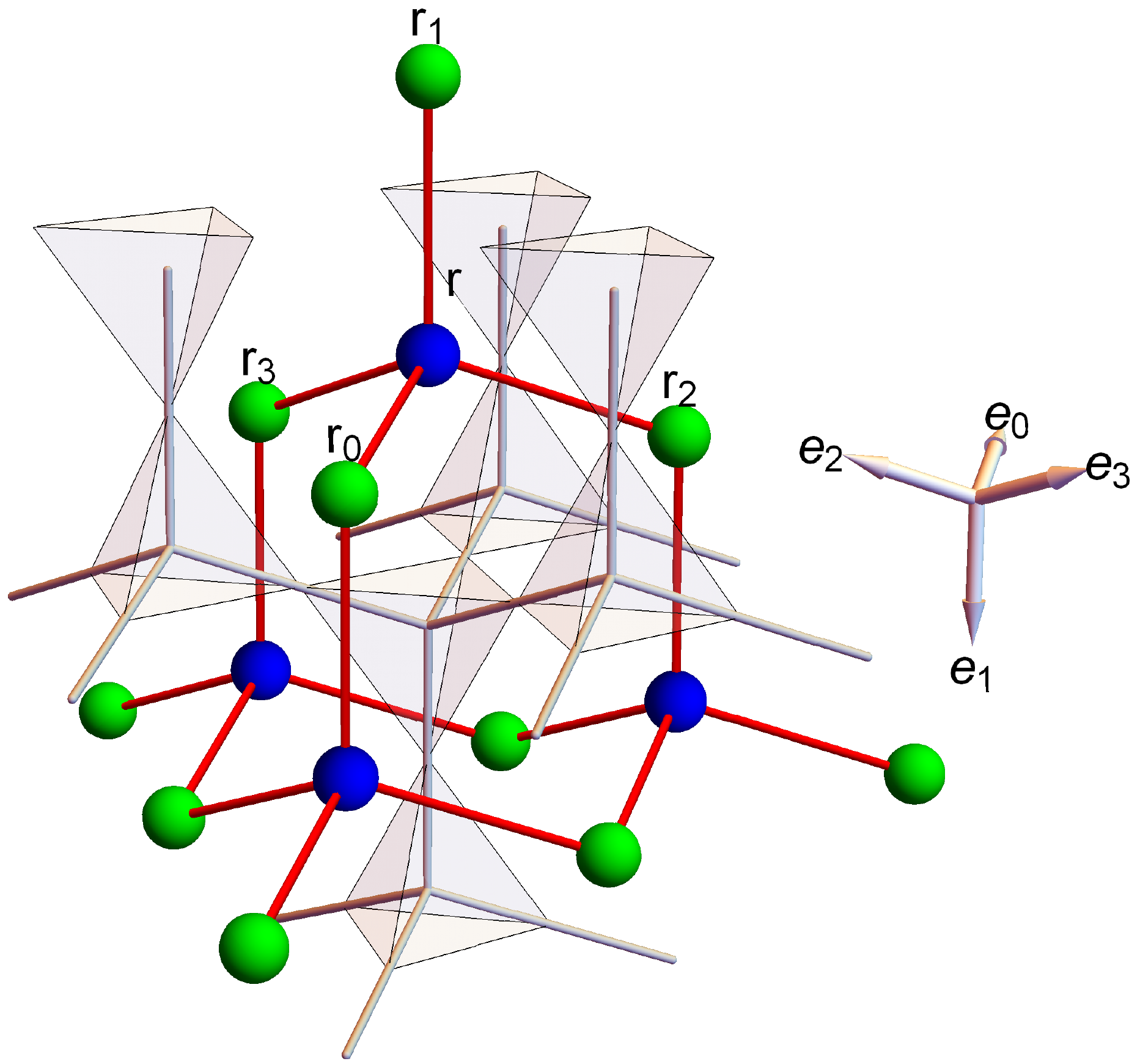}
\caption{(Color online.) Diamond lattice (in gray line) and the dual diamond lattice 
(in red line). The physical spin is located in the middle of the link on the diamond 
lattice. The diamond lattice is formed by the tetrahedral centers of the original 
pyrochlore lattice. The spinons (``magnetic monopoles'') hop on the diamond (dual diamond) 
lattice. The colored balls correspond to the position of  ``magnetic monopoles.''}
\label{2}
\end{figure}

The pyrochlore U(1) QSL for the effective spin-1/2 moments can 
be accessed by a ring exchange model~\cite{PhysRevB.69.064404}
\begin{equation}
\begin{aligned}
{\cal H}_{\rm ring}= & -\frac{K}{2} \sum_{\hexagon_p} 
(\tau_1^{+}\tau_2^{-}\tau_3^{+}\tau_4^{-}\tau_5^{+}\tau_6^{-} + {\rm H.c.} ) 
\\
\end{aligned}
\label{ring}
\end{equation}
where $K$ is a renormalized energy scale for the low-energy effective theory. 
Here the spin operators are ${\tau^{\pm}_i=\tau^{x}_i \pm i\tau^y_i }$.  
A $z$-direction is defined locally along the $\langle 111 \rangle$-direction
of each site. An elementary hexagonal ring ``$\hexagon_p$'' is formed by six  
neighboring sites ${i=1,...,6}$ on the pyrochlore lattice, and the subindex 
``$p$'' refers to the pyrochlore lattice. One can transform the ring 
exchange model into a compact U(1) lattice gauge theory 
(LGT)~\cite{PhysRevB.69.064404, PhysRevLett.108.037202},
\begin{equation}
\begin{aligned}
{\cal H}_{\rm LGT} = & -K  \sum_{{\hexagon_d}} 
\cos\big[ {\rm curl} A \big] 
+ \frac{U}{2}\sum_{rr^\prime} \big(E_{rr^\prime}-\frac{\epsilon_r}{2} \big)^2 
\\
\end{aligned}
\label{LGT}
\end{equation}
by introducing a pair of lattice gauge fields, i.e., electric 
field ${E_{rr^\prime}=\tau_i^z +1/2}$ and vector gauge potential 
${e^{\pm iA_{rr^\prime}}=\tau_i^{\pm}}$. 
These fields are defined on the nearest-neighbor diamond links $rr^\prime$,
where $r$ and $r'$ are used to label the diamond lattice sites. 
The pyrochlore site $i$ sits at the mid-point of the corresponding link $rr^\prime$. 
Two distinct sublattices $r({\rm \in \text{I}}), r^\prime({\rm \in \RNum{2}})$ 
reside at the centers of two corner sharing tetrahedra of the pyrochlore
lattice. $E_{rr^\prime}$ (integer-valued) and $A_{rr^\prime}$ 
($2\pi$-periodic) form a pair of conjugated fields satisfying 
${[E_{rr^\prime},A_{r_1r_1^\prime}]=i \delta_{rr_1,r^\prime r_1^\prime}}$. 
The lattice curl is defined as summation over all bonds of a diamond hexagon 
${{\rm curl} \, A = \sum_{rr^\prime \in { \hexagon_d} } A_{rr^\prime}}$. 
Here ``$\hexagon_d$'' refers to the elementary hexagon on the diamond 
lattice formed by the tetrahedral centers of the pyrochlore lattice. 
Additionally, an electric field stiffness $U$ term is added,
where ${\epsilon_{r}=+1\,(-1)}, {r\in {\rm \RNum{1}}\,({\rm \RNum{2}})}$. 
In the large $U$ limit, the Hilbert space of the LGT is properly 
casted back to the microscopic spin-1/2 local moment. In the low 
energy and long distance limit, the actual $U$ is renormalized 
compared to the original lattice level.

``Magnetic monopole'' is the topological defect of emergent U(1) gauge 
potential, and is the source and sink of the internal magnetic fields. 
Unlike the spinons that reside on the tetrahedral centers
of the pyrochlore lattice (or the diamond lattice sites), the 
``magnetic monopoles'' live on the dual diamond lattice. 
In the above electric field and gauge field representation, 
the ``magnetic monopole'' variable is not explicit. An 
electromagnetic duality transformation is performed on the  
LGT to expose this variable~\cite{PhysRevB.69.064404,PhysRevB.96.195127}.
Although this is covered in the literature extensively, 
some steps of the derivation are not mathematically straightforward.
We carry out the duality transformation in Appendices~\ref{AppSec1} 
and \ref{AppSec2}, where a special care has been taken for the diamond 
lattice structure. The expression of the dual Hamiltonian is presented here,
\begin{eqnarray}
{\cal H}_{\rm dual}[\theta, a,B] &= & \sum_{\langle r r^\prime \rangle} 
\frac{U}{2}  ( {\rm curl} a_{rr^\prime} - 
\bar{E}_{  rr^\prime  })^2 - \sum_{ \langle \textsf{r} \textsf{r}^\prime \rangle} 
K \cos B_{\textsf{r} \textsf{r}^\prime}  
\nonumber \\
& & - t \sum_{\textsf{r} \textsf{r}^\prime} \cos(\theta_{\textsf{r}} 
  - \theta_{\textsf{r}^\prime}+ 2\pi a_{\textsf{r} \textsf{r}^\prime} )  
\end{eqnarray}
where $\textsf{r}, \textsf{r}^\prime$ represent dual diamond lattice 
sites as plotted in Fig.~\ref{2}. A rotor variable $e^{\pm i\theta_{\textsf{r}}}$ 
is proven to be the creation/annihilation operator of the ``magnetic monopole'' 
(see Appendix~\ref{AppSec2}). 
We restore the bosonic nature of the ``magnetic monopole'' variable by introducing
$\Phi_\textsf{r}\equiv \rho_\textsf{r} e^{i\theta_\textsf{r}}$,
where a unimodular condition $|\Phi_\textsf{r}|= 1$ is often imposed
if one abandons the heavier amplitude fluctuations. 
We arrive at the dual Hamiltonian presented in Eq.~\eqref{dual_ham}.
The dual theory describes the ``magnetic monopole''  
$\Phi_{\textsf{r}}$ hopping on the dual diamond lattice and minimally coupled 
to a dual U(1) gauge field. The dual U(1) gauge field $a_{\textsf{r}\textsf{r}^\prime}$ 
(real-valued) and magnetic field $B_{\textsf{r}\textsf{r}^\prime}$ ($2\pi$-periodic) 
are defined on the link $\textsf{r} \textsf{r}^\prime$ of the dual diamond lattice.
These dual fields are related to the field in the original representation by,
\begin{equation}
\begin{aligned}
& {\rm curl} \, a \equiv \sum_{\textsf{r} \textsf{r}^\prime \in \hexagon_d^\ast } 
a_{\textsf{r} \textsf{r}^\prime}= E_{rr^\prime} - E^0_{rr^\prime} , \\
& B_{\textsf{r} \textsf{r}^\prime} = {\rm curl} \, A 
\equiv \sum_{rr^\prime \in {\hexagon_d} } A_{rr^\prime},  \\
\end{aligned}
\label{a_B}
\end{equation}
where the dual hexagonal ring is labelled by $\hexagon_d^\ast$.
The dual lattice curl is defined as summation over all bonds
of a dual hexagon. The definitions in Eq.~\eqref{a_B} guarantee 
that the commutation relation is satisfied 
${[ B_{\textsf{r}\textsf{r}^\prime},  
a_{\textsf{r}_1 \textsf{r}^\prime_1} ] 
= i \delta_{\textsf{r}\textsf{r}_1,\textsf{r}^\prime\textsf{r}^\prime_1} }$. 
A background electric field $E^0_{rr^\prime}$ is introduced in Eq.~\eqref{a_B}
to ensure the lattice curl of dual gauge field is divergenceless. 
Without loss of generality, we choose a 
specific 2-in-2-out spin-ice configuration
for the background electric field, e.g.,
\begin{equation}
\begin{aligned}
E^0_{r,r+\epsilon_r e_0} = & E^0_{r,r+\epsilon_r e_1} =\epsilon_r ,\\ 
E^0_{r,r+\epsilon_r e_2} = & E^0_{r,r+\epsilon_r e_3} =0 . \\ 
\end{aligned}
\label{E_0_main}
\end{equation}
For the future reference, we define another electric field composed 
of the background electric field and an offset field,
\begin{equation}
\begin{aligned}
\bar{E}_{rr^\prime} = & E^0_{rr^\prime} - \frac{\epsilon_{r}}{2} . \\
\end{aligned}
\label{E_bar}
\end{equation}

\section{Induction of dual U(1) gauge flux by Zeeman coupling}
\label{sec3}

The pyrochlore U(1) QSL is in the deconfined phase of the 3+1D LGT. 
It supports both deconfined spinons and deconfined ``magnetic monopoles,'' 
as well as the gapless U(1) gauge photon~\cite{PhysRevB.69.064404} 
(see Table~\ref{notation}). In the inelastic neutron scattering 
experiments, these executions correspond to the continuous 
excitations in the spectrum. The content of the continuum is actually 
related to the nature of the local moments, which is elucidated 
in Refs.~\onlinecite{PhysRevB.96.195127,Chen1902.07075}.
The $\tau^z$-$\tau^z$ correlation contains the information 
of both the (gapped) ``magnetic monopole'' continuum 
and the (gapless) gauge photon~\cite{PhysRevB.96.195127}. 
Moreover, the spectral structure of the continuum is intimately 
tied to the symmetry fractionalization of the spinons and ``magnetic 
monopoles''~\cite{PhysRevB.96.195127,Chen1902.07075,PhysRevB.96.085136}. 
Although these results are quite useful, they are all 
consequences of the deconfinement and fractionalization, 
not a direct evidence of the matter-gauge coupling. 
To demonstrate the consequence of the matter-gauge coupling,
let us consider the Landau level physics in the system of electrons.
The Coulomb interaction between the electrons 
is the consequence of the facts that the electron carries the
U(1) gauge charge and the photon mediates the 
interaction through the electron-photon coupling. 
The electron-gauge coupling of the electrons can be revealed
 through the quantum oscillation of a metal
in external magnetic fields, which arises from the population
of electronic Landau levels . In our case, 
the ``magnetic monopole'' is coupled to the internal dual 
U(1) gauge field, and the ``magnetic monopole'' is bosonic
and gapped. So there does not exist the usual quantum oscillation. 
Moreover, the internal U(1) gauge flux is not obviously tunable. 
Our key observation is that the external field could 
generate an internal dual U(1) gauge flux for the 
``magnetic monopoles.'' This is already pointed in Sec.~\ref{sec1}.
In the following, we embark on explaining this point with 
the non-Kramers doublets.

For the non-Kramers doublets, only the local $z$-component 
of the effective spin is odd under the time reversal symmetry. 
The Zeeman coupling of the effective spin to the external field 
is given as 
\begin{equation}
\begin{aligned}
{\cal H}_{\rm Zeeman}= & - H_0 \sum_{i} ({\hat n} \cdot {\hat z}_i) \tau_i^z \\
\simeq & - H_0 \sum_{\langle rr^\prime \rangle}  ({\hat n} \cdot {\hat z}_i)
({\rm curl}\ a_{rr^\prime}-\bar{E}_{rr^\prime}), \\
\end{aligned}
\end{equation}
where the first line is written with the microscopic spin language
while the second line is expressed in terms of the emergent variables
in the pyrochlore U(1) QSL phase. Here the link $\langle rr^\prime \rangle$
on the diamond lattice is identical to the pyrochlore lattice site $i$, 
$\hat{n}$ defines the direction of the magnetic field,
and ${\hat z}_i$ denotes the local $z$-direction of on the lattice site $i$.
A weak external magnetic field polarizes the spins in each pyrochlore tetrahedron partially,
and throughout we work in the weak field regime such that the 
U(1) QSL state is preserved, namely the lattice gauge theory is in its deconfined phase.  
Hence, the ``magnetic monopole'' representation in Eq.~\eqref{dual_ham} 
remains to be a valid picture for the system.

The Zeeman coupling term enters into the dual Hamiltonian Eq.~\eqref{dual_ham} 
as a modification of the background electric field distribution,
\begin{equation}
\begin{aligned}
{\cal H}_{\rm dual}(H_0)= &  \sum_{\langle rr^\prime \rangle} 
                             \frac{U}{2} ({\rm curl}\ a_{rr^\prime}-\bar{E}^\prime_{rr^\prime})^2 - \cdots  \\
& \bar{E}^\prime_{rr^\prime} = \bar{E}_{rr^\prime} + \frac{H_0}{U}({\hat n} \cdot {\hat z}_i) . \\
\end{aligned}
\end{equation}
We observe that the external field modifies the internal 
dual U(1) gauge flux and thereby generates an emergent Lorentz 
force on the ``magnetic monopoles.''  The motion of the 
``magnetic monopoles'' will be twisted by the induced 
dual U(1) gauge flux, giving rise to the TTHE of ``magnetic 
monopoles.'' This is a direct manifestation and unbiased 
signature of the emergent ``monopole''-gauge coupling. 
This phenomenon serves as an analog of the Lorentz force 
for the electron motion on the lattice, except that the 
Lorentz force here is emergent and arises from the induction of 
the internal dual U(1) gauge flux via the Zeeman coupling.

The Zeeman coupling depends sensitively on the local crystal field axis.
Thus, the induced dual U(1) gauge flux depends on the lattice geometry 
and the field orientation, i.e., the mean field value of dual gauge flux $\langle {\rm curl} \ a \rangle $ 
is related to the induced local magnetization $\langle \tau^z \rangle$. 
Without the Zeeman field, the dual U(1) gauge flux is $\pi$ for the 
elementary hexagon on the dual diamond lattice. The Zeeman coupling 
breaks the time reversal symmetry 
and shifts the dual U(1) gauge flux from $\pi$ by a finite portion
\begin{equation}
\begin{aligned}
 2\pi \langle {\rm curl}\ a_{rr^\prime} \rangle 
=& \pi + 2\pi \frac{H_0}{U} ( {\hat n} \cdot {\hat z}_i ) \quad {\rm mod}\,(2\pi), \\
\end{aligned}
\label{MF_a}
\end{equation}
where $\langle {\rm curl}\ a_{rr^\prime} \rangle$ represents 
a mean-field solution for the dual gauge flux. 
The parameter $U$ is often unknown. 
Physically, the induced flux can be obtained from the induced local 
magnetization that is given as,
\begin{equation}
 \langle \tau^z_i \rangle \equiv \chi_i  (\hat{n} \cdot \hat{z}_i) H_0 ,
\end{equation}
which depends on the local spin susceptibility $\chi_i$ along the $z$-direction on each site $i$.
In the weak field limit, $\chi_i$ should be uniform by definition and symmetry requirement.
It is also a constant due to the strong spin-orbit coupling in the system. 
The above equations give us the relations between the induced dual U(1) flux and 
the physical magnetization. 

With the mean-field solution of dual U(1) gauge flux in the presence of the Zeeman field,
we write down a mean-field Hamiltonian for the ``magnetic monopoles,''
\begin{equation}
{\cal H}_{\rm MF} =  -\frac{t}{2} \sum_{\textsf{r}\textsf{r}^\prime} 
e^{-i2\pi  a^0_{\textsf{r}\textsf{r}^\prime }}
\Phi^\dagger_{\textsf{r}^\prime} \Phi_{\textsf{r}}^{} + {\rm H.c.} - 
\mu \sum_{\textsf{r}}\Phi^\dagger_{\textsf{r}} \Phi_{\textsf{r}}^{} , 
\label{MF_ham}
\end{equation}
where $a^0_{\textsf{r}\textsf{r}^\prime}$ represents a gauge choice for the dual 
U(1) gauge field. The dual gauge field is fixed at a particular mean-field solution,
and its conjugate field, namely the internal magnetic field, is omitted.
Therefore, the Hamiltonian in Eq.~\eqref{MF_ham} describes the hopping of 
``magnetic monopoles'' in the presence of a dual U(1) gauge field, whose 
fluctuation has been ignored.

\section{Topological thermal Hall effect}
\label{sec4}

In the previous sections, we have explained our ideas and the physical 
origin of the TTHE for the ``magnetic monopoles.'' 
Here we further establish the theoretical framework to 
demonstrate the TTHE and make specific predictions for the experiments.

\subsection{General framework}
\label{sec4a}

To extract information out of the twisted motion of the ``magnetic monopoles,'' 
we perturb the system with a temperature gradient in the plane perpendicular 
to the external magnetic field. 
In the standard linear response theory, the small external perturbation 
appears in the Hamiltonian. The effect of the temperature gradient 
$T({\bf r})\simeq T_0[1-\psi({\bf r})]$ takes place in the Boltzmann factor,  i.e.,
$ e^{-{\cal H}/k_{\rm B}T({\bf r})} \simeq e^{-[1+\psi({\bf r})]{\cal H}/k_{\rm B}T_0} $.
Theoretical framework tackling with this problem has been
proposed by Luttinger~\cite{PhysRev.135.A1505}.
By coupling the Hamiltonian with a pseudo-gravitational potential $\psi({\bf r})$,
they are able to incorporate the temperature gradient 
into a perturbed Hamiltonian $\bar{\cal H}({\bf r})=[1+\psi({\bf r})]{\cal H}$.

We start from the mean-field Hamiltonian in Eq.~\eqref{MF_ham},
and treat the dual diamond lattice structure carefully.
The pseudo-gravitational potential $\psi_{\textsf{r}}$ couples with 
an energy density operator ${\cal H}_{\textsf{r}}$.
The coupling is turned on for one type of the dual sites
with 
\begin{equation}
\begin{aligned}
\bar{\cal H} = & \sum_{\textsf{r} \in 
{\rm \RNum{1}}} (1+\psi_{\textsf{r}}) {\cal H}_{\textsf{r}} , 
\\
\end{aligned}
\end{equation}
The energy density operator at a dual site $\textsf{r}$ is defined as 
\begin{equation}
{\cal H}_{\textsf{r}} = - \frac{t}{2}  \sum_{\textsf{r}^\prime \in \textsf{r}} 
 e^{-i2\pi a^0_{\textsf{r}\textsf{r}^\prime}} 
\Phi_{\textsf{r}^\prime}^\dagger \Phi_{\textsf{r}}^{} 
+ {\rm H.c.},
\end{equation}
where the summation is over four nearest neighbor dual sites 
$\textsf{r}^\prime \in \textsf{r}$, which are labelled in Fig.~\ref{2}.
The chemical potential term is omitted in the energy density operator,
since it has no contribution to the transport properties below.
The energy density is not modified upon the addition of pseudo-gravitational potential, 
since the four nearest neighbors necessarily belong to the type-{\rm \RNum{2}} sites.
We work through the lattice version of continuity equation 
for the energy density operator, 
\begin{equation}
\dot{\cal H}_{\textsf{r}}+ \sum_{\textsf{r}^\prime \in \textsf{r}} 
{\cal J}^{E}_{\textsf{r}\textsf{r}^\prime}=0.
\label{continuity}
\end{equation}
Working through the above continuity equation with the modified local Hamiltonian
$(1+\psi_{\textsf{r}}) {\cal H}_{\textsf{r}}$, we obtain the modified energy 
current operator~\cite{PhysRevB.91.125413, PhysRevB.89.054420},
\begin{equation}
{\cal J}^{E}_{\textsf{r}\textsf{r}^\prime}=(1+\psi_{\textsf{r}^\prime}) {\cal J}^{0,E}_{\textsf{r}\textsf{r}^\prime},
\end{equation}
where ${\cal J}^{0,E}_{\textsf{r}\textsf{r}^\prime}$ represents 
the original energy current, which has a form
\begin{equation}
\begin{aligned}
{\cal J}^{0,E}_{\textsf{r}\textsf{r}^\prime}
= &  \frac{t^2}{2} \sum_{ \textsf{r}_1 \in \textsf{r}^\prime} i \Phi_{\textsf{r}}^\dagger \Phi_{\textsf{r}_1} e^{i2\pi (a^0_{\textsf{r}\textsf{r}^\prime}+a^0_{\textsf{r}^\prime\textsf{r}_1})}  + {\rm H.c.} .\\
\end{aligned}
\label{J_0}
\end{equation}

Under the choice of a uniform potential gradient~\cite{PhysRevB.91.125413}, 
we have ${\psi_{\textsf{r}}={\bf r}_{\bf i} \cdot \nabla \psi}$,
where ${\bf r}_{\bf i}$ represents the position of a unit cell ${\bf i}$.
The dual lattice links constituting the unit cell ${\bf i}$ 
are labeled as ${\textsf{r}\textsf{r}^\prime \in {\bf i} }$.
The choice of this unit cell depends on the dual gauge fixing condition, 
which is specified in Sec.~\ref{sec4b}.
In terms of the unit cell coordinate ${\bf r}_{\bf i}$, 
we rewrite the modified energy current operator as~\cite{PhysRevB.91.125413}
\begin{equation}
\begin{aligned}
J_\alpha^E({\bf i}) = & J_{\alpha}^{0,E}({\bf i})  + J_{\alpha}^{1,E}({\bf i}) ,\\ 
J_\alpha^{1,E}({\bf i}) = &  \big[J_\alpha^{0,E}({\bf i}) r^\beta_{\bf i} \big] \nabla_\beta \psi , \\
\end{aligned}
\end{equation}
where $\alpha,\beta = x,y,z$. The energy density vector 
at the unit cell ${\bf i}$ is defined as,
\begin{equation}
J_\alpha^{0,E}({\bf i}) = 
\sum_{\textsf{r},\textsf{r}+\epsilon_{\textsf{r}} e_\mu \in {\bf i} } 
( \epsilon_{\textsf{r}} e_\mu \cdot \hat{\alpha} ) 
{\cal J}^{0,E}_{\textsf{r},\textsf{r}+\epsilon_{\textsf{r}} e_\mu} .
\end{equation}

The linear response of the pseudo-gravitational field enters 
into the energy current expectation value in a two-fold way.
Besides the contribution from the distribution function~\cite{PhysRevLett.104.066403},
there is an additional contribution from the current operator.
At the linear order in $\nabla \psi$, we have
\begin{equation}
\langle J_\alpha^{E} \rangle = {\rm Tr}[\rho_0 J_\alpha^{1,E}]
+ {\rm Tr}[\rho_1 J_\alpha^{0,E}],
\label{J_E_main}
\end{equation}
where $\rho_0$ is the equilibrium distribution function and
$\rho_1$ is a first order perturbed distribution function.
A statistical force from the temperature gradient 
is equivalent to a dynamical force induced by pseudo-gravitational potential.
The dynamical force acts on the ``magnetic monopole'' affecting its motion.
By counting all the contributions due to the temperature gradient 
at the first order, the thermal Hall coefficient is calculated and 
has an expression~\cite{PhysRevLett.106.197202, PhysRevB.89.054420}
\begin{equation}
\begin{aligned}
\kappa_{xy} = & -\frac{k_{\rm B}^2T}{N^3} \sum_{{\bf k}} 
\sum_{n=1}^{6}\Big\{ c_2\big[g(E_{n,{\bf k}})\big]
- \frac{\pi^2}{3}\Big\} \Omega_{n,{\bf k}},  \\
\end{aligned}
\label{k_xy}
\end{equation}
where $c_2(x)=(1+x)[\ln(1+x)/x]^2-(\ln x)^2- {\rm Li}_2(-x)$, 
and ${\rm Li}_2(x)$ is a polylogarithmic with ${n=2}$, or the dilogarithm function.
Here ${g(\epsilon)=[e^{\epsilon/k_{\rm B}T}-1]^{-1}}$ is the Bose distribution function.
$E_{n,{\bf k}}$ is the eigen-energy of the ``monopole'' Hamiltonian for the $n$'th band
at the momentum space ${\bf k}$-point. 
Here, the Berry curvature and Chern number for the $n$'th band are defined as,
\begin{equation}
\begin{aligned}
&\Omega_{n,{\bf k}}= i \langle \partial_{k_x} u_{n,{\bf k}}|\partial_{k_y} u_{n,{\bf k}}\rangle +{\rm c.c.} ,\\
&{\cal C}_n(k_z) = \frac{1}{2\pi} \int_{\rm BZ} dk_x dk_y\ \Omega_{n,{\bf k}}, \\
 \end{aligned}
\label{Berry_Chn}
\end{equation}
where $|u_{n,{\bf k}}\rangle$ is the periodic part of the Bloch wave 
function for the $n$'th band at ${{\bf k}=(k_x,k_y,k_z)}$.
The formula indeed shows that the thermal Hall current is generated
by the Berry curvature of the ``monopole'' bands.
Due to the time reversal symmetry breaking in the presence of the 
gauge flux, we can have non-vanishing distribution of Berry curvatures,
that gives rise to a finite thermal Hall coefficient.

For our purpose, it is sufficient to consider the TTHE in the 
presence of the mean-field dual U(1) gauge flux.  
At the mean-field level, the dual U(1) gauge field is fixed by the 
background electric field, and the internal magnetic field is absent.
The energy current in Eq.~\eqref{J_0} is obtained 
by using the Hamiltonian in Eq.~\eqref{MF_ham}.
Beyond the mean-field solution, we find that 
the gauge fluctuations give the thermal current operator a correction.
The expression and derivation of this additional contribution 
is presented in Appendix~\ref{AppSec3}. The (gapless) gauge 
photon contributes directly to the thermal conductivity $\kappa_{xx}$ 
around the same energy scale as the ``magnetic monopoles'' 
except that it remains active down to the lowest energy/temperature
and the contribution can directly come from the (fluctuating) Maxwell term.
In addition, the spinons would contribute to the thermal effect $\kappa_{xx}$
when the temperature is relatively high to activate spinons. 
In our current theoretical understanding, the ``magnetic monopoles''
are singled out to be responsible for the thermal Hall conductivity,
and the TTHE in this work refers particularly to the ``magnetic monopole''
thermal Hall effect.

\begin{figure}[htbp]
\centering
\includegraphics[width=7cm]{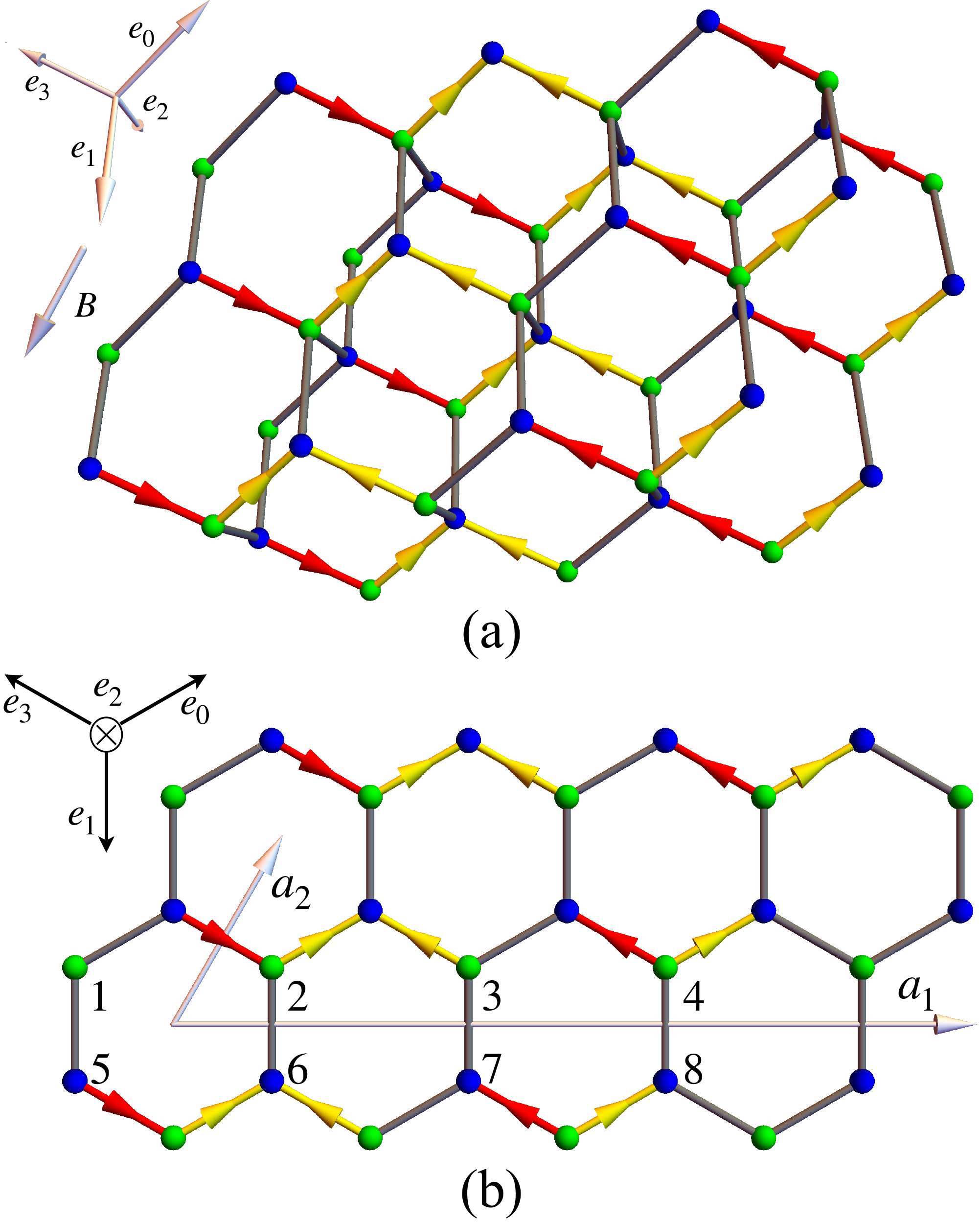}
\caption{(Color online.) (a) Gauge fixing on the dual diamond lattice. 
The yellow (red) arrow on the dual links represents a finite phase $\phi=\pi\ (\pi/2)$
picked up by ``magnetic monopole'' when hopping along the pointed direction. While, 
the ``monopole'' hopping on the gray bond is free of the phase, {\sl i.e.} ${\phi=0}$.
(b) A projected view of lattice in the plane perpendicular to $e_2$-direction. 
The dual sites with indexes $1,2,...,8$
constitute a magnetic unit cell. The basis vectors within the plane are labelled 
by $a_1, a_2$.}
\label{5}
\end{figure}

\subsection{The modified ``magnetic monopole'' bands under the magnetic field}
\label{sec4b}

To demonstrate the TTHE for the ``magnetic monopoles'' in the pyrochlore U(1) QSL,   
we first evaluate the ``magnetic monopole'' band structure under the magnetic field. 
Generically speaking, when a generic magnetic field is applied, the ``magnetic monopole''
should develop a Hofstadter band structure as the induced flux is incommensurate. 
The corresponding continuum of ``magnetic monopoles'' in the $\tau^z$-$\tau^z$ 
correlation is converted into the continuum from the monopole Hofstadter band. 
It would be interesting to search for this evolution in
the inelastic neutron scattering measurements.

We choose the external field to be aligned in the direction
${{\hat n}=\langle 0\bar{1}\bar{1} \rangle}$. 
To proceed with the mean-field Hamiltonian in Eq.~\eqref{MF_ham},
we fix the Zeeman coupling strength such that 
the dual U(1) gauge flux is commensurate with the lattice. 
A convenient case is considered here with $H_0/U= \sqrt{6}/8$,
so that we have
\begin{equation}
2\pi\ {\rm curl} \, a^0_{r,r+ e_\mu}=
\left\{ \begin{array}{cc}
\pi - \pi/2, & \ \ \ \ \ \mu=0,\\
\pi + \pi/2, & \ \ \ \ \ \mu=1,\\
\pi +0, & \ \ \ \ \ \mu=2,\\
\pi +0, & \ \ \ \ \ \mu=3.\\
\end{array}\right.
\label{a_1}
\end{equation}
where $a^0_{\textsf{r}\textsf{r}^\prime}$ represents a gauge choice for the dual U(1) 
gauge field. The gauge fixing condition on the three-dimensional(3D) dual diamond lattice 
is illustrated in Fig.~\ref{5}(a). The yellow (red) arrow on the dual links indicates 
that a ``magnetic monopole'' picks up a finite phase $\phi=\pi\ (\pi/2)$
while hopping along the pointed direction. Gray links have zero phases $\phi=0$.
The gauge fixing condition is expressed as,  
\begin{equation}
\begin{aligned}
2\pi a^0_{\textsf{r},\textsf{r} + e_\mu} = &  \xi_\mu ({\bf q}_1\cdot \textsf{r}) + \eta_\mu ({\bf q}_2\cdot \textsf{r}), \ \ \ \textsf{r}\in {\rm \RNum{1}}, \\
{\bf q}_1= & 2\pi (100), \   \xi_\mu=  (1001), \\
{\bf q}_2= & \pi (100), \ \eta_\mu=(0001). \\
\end{aligned}
\label{a_2}
\end{equation}
where the ``monopole'' charge is assumed to be unit $q_m =1$. 
With this gauge choice, gauge fields are non-vanishing 
at the links locating within a quasi-2D plane perpendicular 
to the $e_2$-direction. The dual lattice hexagons form a 
honeycomb-like structure in this quasi-2D plane. 
A projected view of the lattice in this plane is illustrated in Fig.~\ref{5}(b).
We define a magnetic unit cell consisting of eight distinct dual diamond sites.
A 3D super-lattice is defined by a new set of primitive vectors 
$a_\nu, \nu=1,2,3$ with 
\begin{equation}
\begin{aligned}
& a_1= 4(e_0-e_3) , \\
& a_2= e_0-e_1 , \\
& a_3= e_3- e_2.\\
\end{aligned}
\end{equation}
where the basis convention of these vectors and the corresponding 
reciprocal basis vectors are given in Appendix~\ref{AppSec4}.    

\begin{figure}[b]
\centering
\includegraphics[width=7cm]{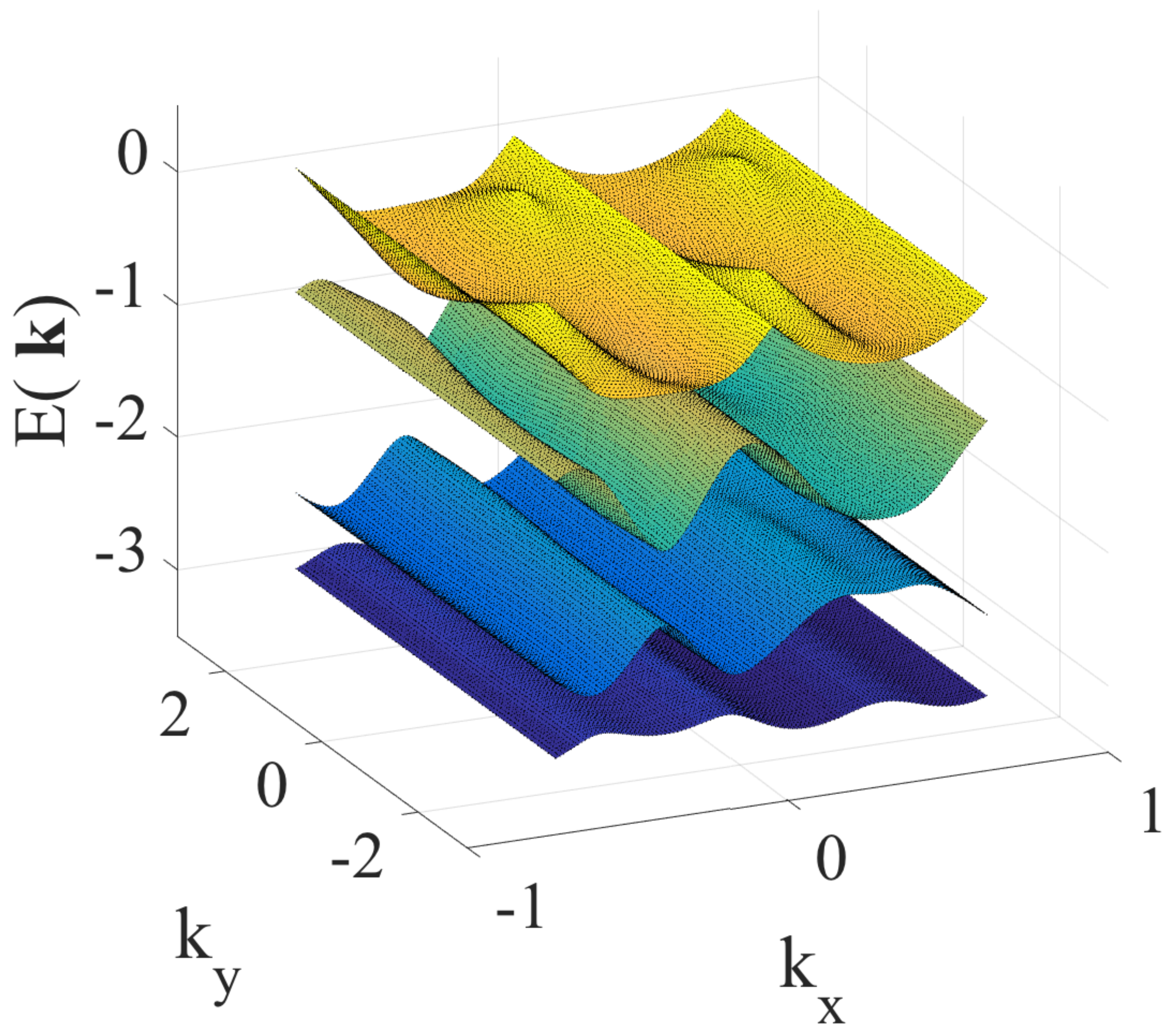}
\caption{(Color online.) 
Energy dispersions of $H_{\rm hop}({\bf k})$ for the four hole bands 
are plotted in the $(k_x, k_y)$ Brillouin zone with a perpendicular 
momentum ${k_z=-3\pi/8}$. The unit of momentum $(k_x, k_y)$ is $a^{-1}$, 
with ``$a$'' being the length of the dual diamond link.
}
\label{3}
\end{figure}

This is a commensurate case so that one can work out from Eq.~\eqref{a_1}.
The ``magnetic monopole'' mean-field Hamiltonian is given by, 
\begin{equation}
 {\cal H}_{\rm MF}({\bf k})= {\cal H}_{\rm hop}({\bf k}) -\mu \, {\bf I}_{8\times 8}
 \label{band}
 \end{equation}
The hopping Hamiltonian takes a particle-hole symmetric form with respect to 
exchanging type-${\rm \RNum{1}}$ and ${\rm \RNum{2}}$ sublattice sites of the dual
diamond lattice, 
\begin{equation}
 {\cal H}_{\rm hop}({\bf k}) =-t \left(\begin{array}{cc} 0 & h({\bf k}) \\ h^\dagger({\bf k}) & 0 \\ \end{array}\right)
\end{equation}
with the hopping Hamiltonian between type-${\rm \RNum{1}}$ and ${\rm \RNum{2}}$ 
sublattice sites given by
\begin{widetext}
\begin{align}
 h({\bf k}) = \left(\begin{array}{ccccc} 
e^{ik_0}+ e^{ik_1} & 0 & 0 & e^{ik_3}+ e^{ik_2} \\
e^{ik_3}e^{i(\pi+\pi/2)}+ e^{ik_2}  & e^{ik_0}e^{i\pi}+ e^{ik_1}  & 0 & 0 \\
0  & e^{ik_3}e^{i\pi}+ e^{ik_2}  & e^{ik_0}+ e^{ik_1}& 0 \\
0 & 0 &  e^{ik_3}e^{i(3\pi/2+\pi)}+ e^{ik_2} &e^{ik_0}e^{i\pi}+ e^{ik_1} \\
\end{array}\right) 
\end{align}
\end{widetext}
where ${{k_\nu \equiv {\bf k}\cdot a_\nu}, \nu=1,2,3.}$
Due to the particle-hole symmetry, the energy spectrum of 
${\cal H}_{\rm hop}({\bf k})$ comes with positive-negative pairs.
The four hole bands are plotted in a ($k_x,k_y$) Brillouin 
zone with a perpendicular momenta ${k_z=-3\pi/8}$ (see Fig.~\ref{3}).
A chemical potential to remedy the negative energy situation 
is added in the mean-field Hamiltonian so that the ``magnetic monopole'' 
remains gapped.

\subsection{Topological thermal Hall effect of ``magnetic monopoles''}
\label{sec4c}

\begin{figure}[htbp]
\centering
\includegraphics[width=8.5cm]{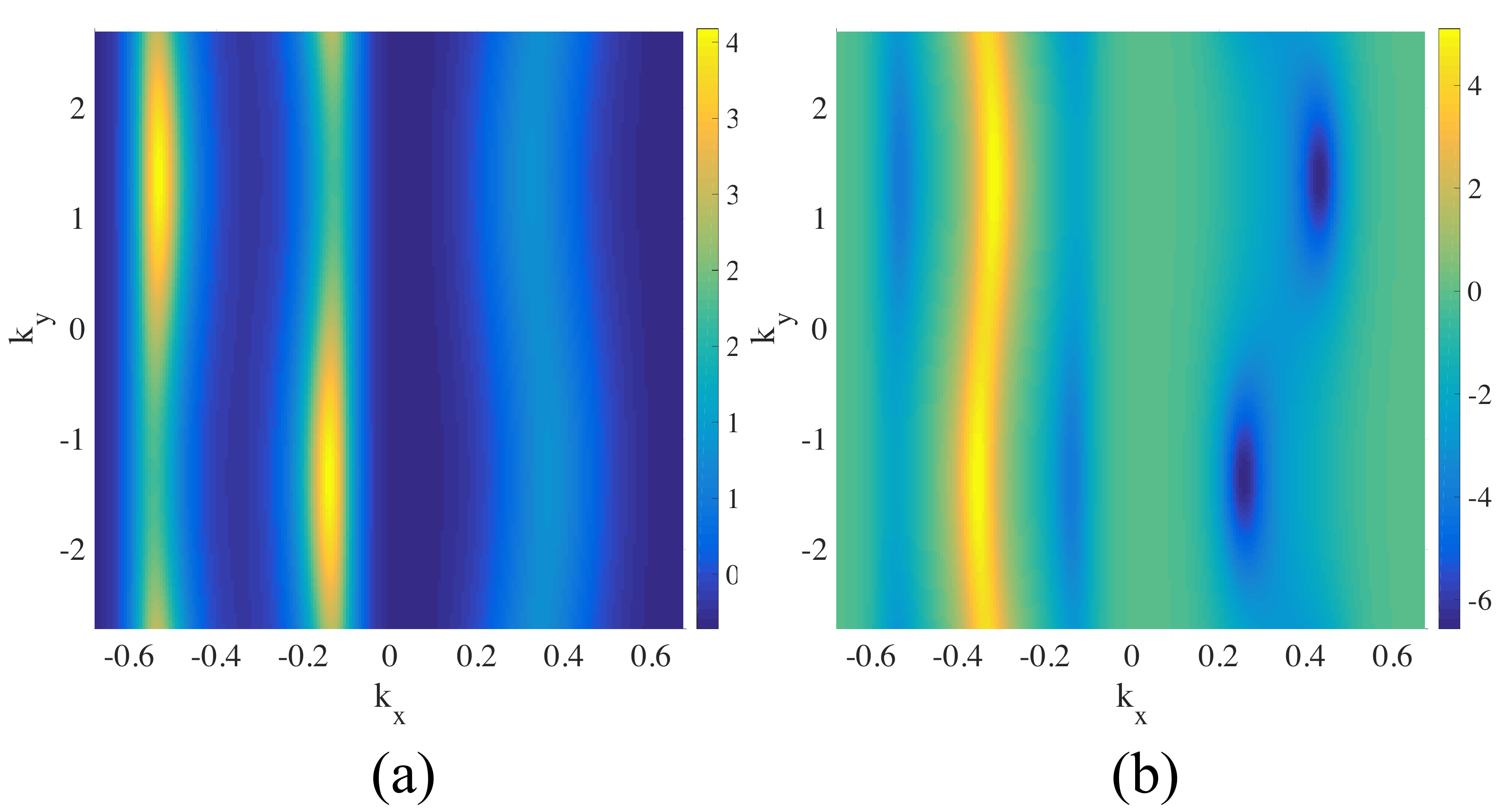}
\caption{(Color online.) Berry curvatures of the lowest two bands in 
the ${(k_x, k_y)}$ Brillouin zone with the perpendicular momentum 
${k_z=-3\pi/8}$. The unit of momentum ${(k_x, k_y)}$ is $a^{-1}$. 
(a) The lowest band ${n=1}$; (b) Second lowest band ${n=2}$.}
\label{4}
\end{figure}

With the above setup and preparation, we here carry out the 
calculation for the TTHE of ``magnetic monopoles''and
show its temperature dependence. First, we evaluate the Berry curvatures
for the ``monopole'' bands. The Berry curvatures of the lowest two
``magnetic monopole'' bands are plotted in the ($k_x$-$k_y$) plane 
with $k_z$ locating at the BZ boundary in Fig.~\ref{4}. 
The Chern number of the lowest band at any given $k_z$ is a positively quantized number ${{\cal C}_1(k_z)=1}$.
The second lowest band is endowed with a non-negative, quantized Chern number, namely ${\cal C}_2(k_z)=-1, 0$.
The two lowest lying bands are of opposite Chern numbers for a majority of $k_z$ points.

\begin{figure}[t]
\centering
\includegraphics[width=8cm]{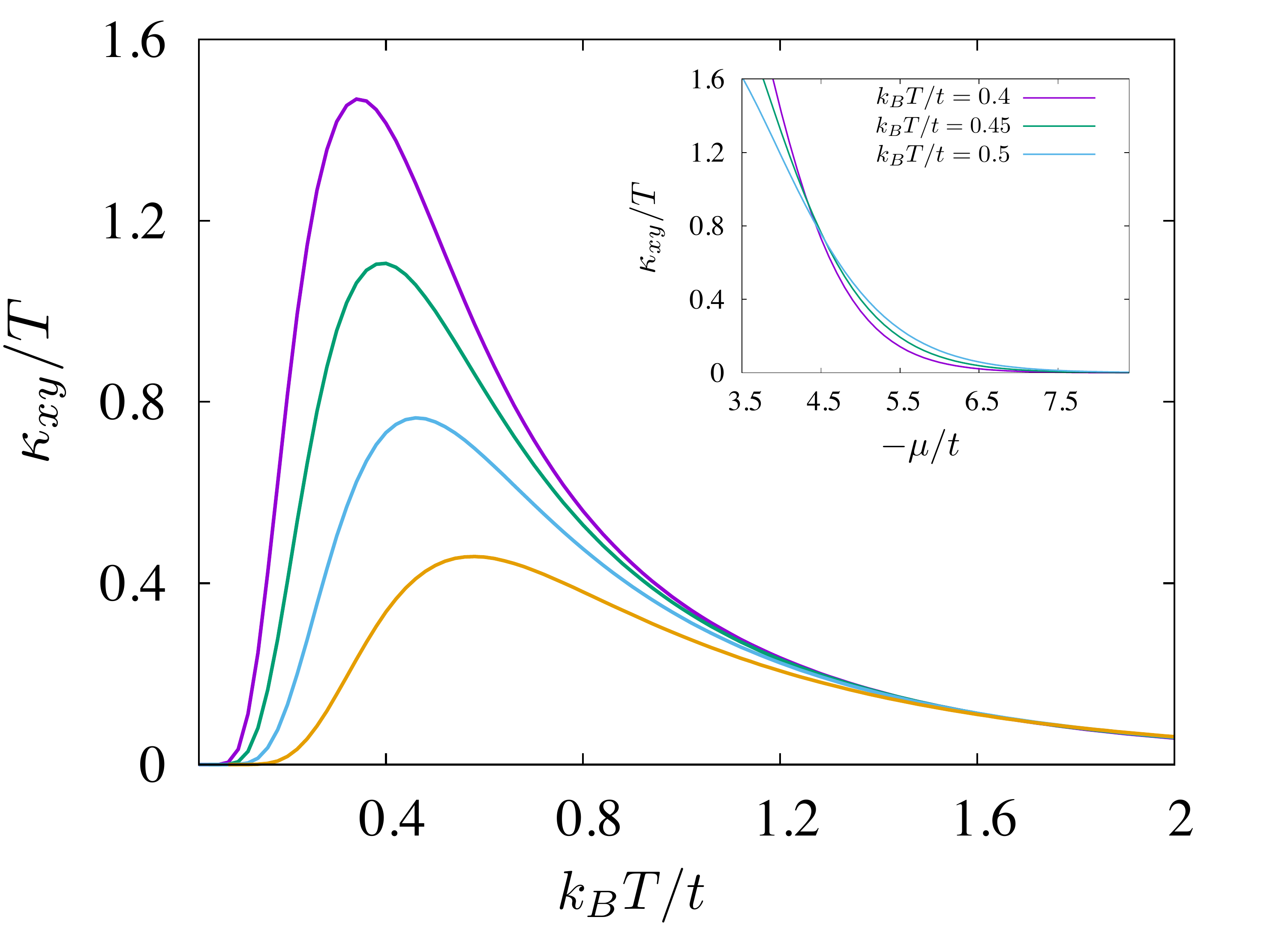}
\caption{(Color online.) The ``magnetic monopole'' thermal Hall coefficient $\kappa_{xy}/T$ 
versus the temperature $k_B T/t$. Curves with different colors (from top to bottom) are plotted 
with a decreasing sequence of chemical potential ${-\mu/t=4, 4.2, 4.5, 5}$.  
The thermal Hall coefficient $\kappa_{xy}/T$ is 
in a unit of $k_B^2/(2\pi \hbar a)\simeq 2.8\times10^{-4}$ W/(K$^2$m).
Inset: The thermal hall coefficient $\kappa_{xy}/T$ is plotted 
versus the chemical potential $-\mu/t$ for a set of temperatures.}
\label{fig6}
\end{figure}

We calculate the thermal Hall coefficient using ``monopole'' bands in Eq.~\eqref{band}. 
The temperature dependence of the thermal Hall coefficient $\kappa_{xy}/T$ is depicted 
in Fig.~\ref{fig6}. With the increasing temperatures, $\kappa_{xy}(T)/T$ grows from zero, 
then, shows a non-monotonic behavior. Eventually, $\kappa_{xy}(T)/T$ drops to 
zero in the high temperature limit. The trend of this curve can be understood 
from Eq.~\eqref{k_xy}, which consists of a product of the Berry curvature and 
a function $c_2$. The function $c_2(g)$ is a monotonically increasing function 
of the occupation $g(\epsilon)$, which has a minimum value ${c_2=0}$ at ${g=0}$ 
and saturates to a maximum value $\pi^2/3$ in the limit $g\rightarrow +\infty$.
In the zero temperature limit, all bands are unoccupied, 
so that the thermal Hall coefficient vanishes. As the temperature increases, 
the lowest band starts to have a finite occupancy, giving rise to the increase 
of $\kappa_{xy}(T)/T$. If we further increase the temperature, the second lowest band, 
with opposite sign of Berry curvature, are activated, which explains the drop 
of the curve. Eventually, all bands are equally populated in the high temperature limit,
although at very high temperatures the ``magnetic monopole'' and U(1) QSL simply 
breaks down. The $\kappa_{xy}(T)/T$ is proportional to the total Chern number, 
which has a vanishing value. Alternatively, one can vary the chemical potential 
while keeping the temperature fixed as shown in the inset of Fig.~\ref{fig6}.
The thermal Hall coefficient decreases along with the chemical potential.
The chemical potential shifts all bands into a higher energy regime.
The occupation of all bands becomes smaller, which is responsible 
for the decrease in $\kappa_{xy}(T)/T$.

Finally, we comment on the temperature dependence of the TTHE that takes
 place along different directions and under different external field strengths.
The thermal Hall coefficient $\kappa_{xz}(T)$ shows exact same temperature 
dependence as $\kappa_{xy}(T)$ with an opposite sign, while $\kappa_{yz}(T)$ 
takes a vanishing value at all temperature. In the next subsection, we 
investigate on the external field strength dependence of the thermal Hall 
coefficient. Conclusively, the dependence of the thermal Hall coefficient 
on the field strengths are qualitatively similar.

\subsection{Topological thermal Hall effect in weak external field limit}
\label{sec4d}

The TTHE is related to the Berry curvature of ``magnetic monopole'' 
bands that arise from the induced dual U(1) gauge flux. Under generic magnetic fields,
the flux is incommensurate, and diagonalizing the monopole Hamiltonian in presence 
of arbitrary gauge flux constitutes a 3D Hofstadter problem~\cite{PhysRevB.14.2239,PhysRevB.96.195127}. 
The incommensurability merely brings some calculational complexity, 
but our formalism should be readily extended over there and the 
calculation can be performed numerically.

\begin{figure*}[htbp]
\centering
\includegraphics[width=17cm]{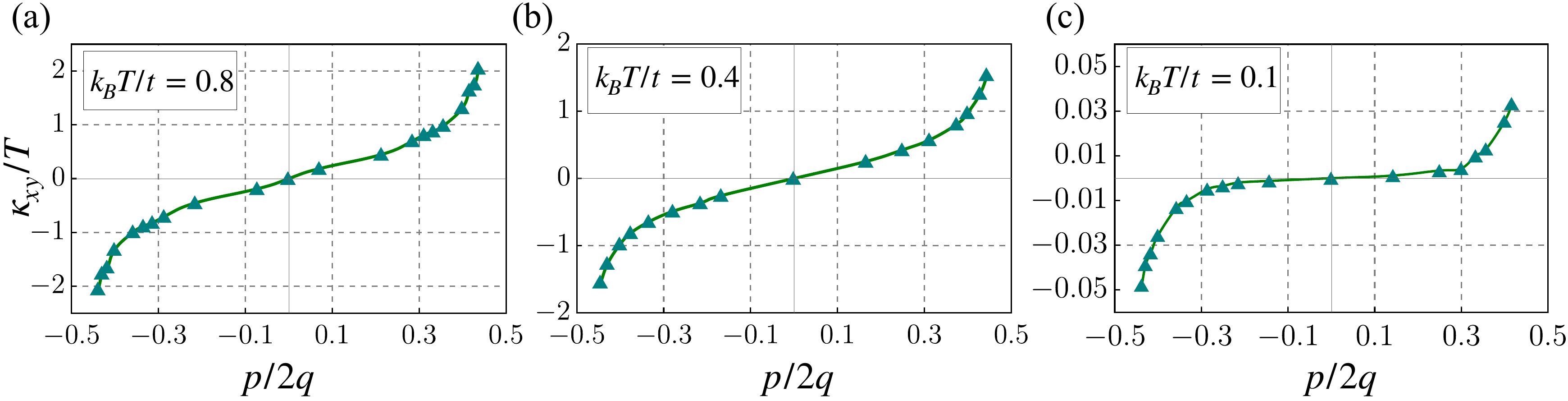}
\caption{The ``magnetic monopole'' thermal Hall coefficient $\kappa_{xy}/T$ 
versus the field strength ratio $p/2q$ for various temperature: 
(a) $k_B T/t=0.8$, (b) $k_B T/t=0.4$, (c) $k_B T/t=0.1$. 
The chemical potential is fixed at $-\mu/t=4$. $\kappa_{xy}/T$ 
is in a unit of $k_B^2/(2\pi \hbar a)\simeq 2.8\times10^{-4}$ W/(K$^2$m).}
\label{7}
\end{figure*}

To demonstrate the usefulness of our theory in the generic commensurate flux cases,
we calculate the thermal Hall coefficient along the same external field direction 
$\hat{n}=\langle 0\bar{1}\bar{1} \rangle$ with the field strength,
\begin{equation}
\begin{aligned}
H_0/U = \frac{\sqrt{3}}{\sqrt{2}} \frac{p}{2q}, \ \ p,q \in {\cal Z}
\end{aligned}
\end{equation}
where $p,q$ are integer numbers, 
and the external field strength is proportional to a gauge flux ratio $p/2q$.
The gauge flux on the diamond hexagon takes a form, 
\begin{equation}
2\pi\ {\rm curl} \, a^0_{r,r+ e_\mu}=
\left\{ \begin{array}{cc}
\pi - (p/q)\pi, & \ \ \ \ \ \mu=0,\\
\pi + (p/q)\pi, & \ \ \ \ \ \mu=1,\\
\pi +0, & \ \ \ \ \ \mu=2,\\
\pi +0, & \ \ \ \ \ \mu=3.\\
\end{array}\right.
\end{equation}
Accordingly, the gauge fixing condition on the dual diamond lattice is given by,
\begin{equation}
\begin{aligned}
2\pi a^0_{\textsf{r},\textsf{r} + e_\mu} = &  \xi_\mu ({\bf q}_1\cdot \textsf{r}) + \eta_\mu ({\bf q}_2\cdot \textsf{r}), \ \ \ \textsf{r}\in {\rm \RNum{1}}, \\
{\bf q}_1= & 2\pi (100), \   \xi_\mu=  (1001), \\
{\bf q}_2= & (p/q)2\pi (100), \ \eta_\mu=(0001). \\
\end{aligned}
\end{equation}
The case we demonstrated in Sec.~\ref{sec4b} is regarded as a special case with $p=1, q=2$.
For the general integer values of $(p,q)$, 
the magnetic unit-cell is enlarged along the $a_2$-direction in Fig.~\ref{5}, 
constituting $4q$ number of distinct dual diamond sites.
With the gauge fixing condition for generic commensurate flux, 
we estimate the TTHE under various external field strength.
We plot the thermal Hall coefficient $\kappa_{xy}/T$ versus 
the gauge flux ratio $p/2q$ in Fig.~\ref{7} 
for three representative temperature points.
The thermal Hall coefficient admits a primitive zone of gauge flux 
ratio $p/2q\in [-0.5,0.5)$, and is periodic with respect to the 
shift of integer gauge flux ratio.
At the zone center and boundary $p/2q =0, -0.5$, 
the TTHE is absent due to the preservation of the time reversal symmetry.
The magnitude of the thermal Hall coefficient $\kappa_{xy}$ increases 
along with the ratio $p/2q \in [0,0.5)$, and changes sign when 
the external field direction is reversed $p/2q \in [-0.5,0)$.
The finite value of thermal Hall coefficient indicates that 
the TTHE is no fluke under the specific gauge choice, rather, 
it is a universal phenomenon in the presence of generic commensurate flux.

For the incommensurate flux case at arbitrary field strength, we can 
approximate the incommensurate flux to a nearby commensurate one 
in the weak field limit. The weak field limit is consistent with 
our previous assumption, which guarantees the existence of the 
underlying QSL ground state. Furthermore, we have considered a 
semiclassical version of the ``monopole'' thermal Hall effect 
under generic magnetic fields and develop a continuous theory 
for this effect. This will be explained in a future work.

\section{Discussion}
\label{sec7}

To summarize, we made the observation from the electromagnetic duality 
that, the external magnetic field could generate emergent electric field 
distribution and thus the dual U(1) gauge flux for the ``magnetic monopoles''.
We developed a formalism to calculate the modulation of the monopole band
structure and the monopole Berry curvature, and explained the physical origin
of the monopole thermal Hall effects.
To provide an illuminating discussion of the implication and underlying insights 
of our results, we first make a comparison between our current theory for the 
pyrochlore U(1) QSL and other U(1) QSLs. Then, we will focus on the 
pyrochlore magnets and make a materials' survey about the thermal Hall 
effects among the pyrochlore U(1) QSL candidate materials.

\subsection{Comparison with other U(1) QSLs in both weak and strong Mott regimes}
 
 Thermal Hall effect was suggested for 
the spinon Fermi surface U(1) QSLs in the weak Mott regime. 
This effect is actually quite natural in the weak Mott 
regime~\cite{PhysRevB.73.155115,PhysRevLett.104.066403}.    
Over there, the concept of spinons are not so distinct from the physical 
electrons due to the weak Mott gap and strong charge fluctuations. 
Physically, this can be understood from the fact that the 
external gauge flux enters into the four-spin ring exchange 
interaction~\cite{PhysRevB.51.1922, PhysRevB.73.155115}. 
From the gauge theory description, the internal U(1) gauge 
flux is locked to the external U(1) gauge flux through the strong 
charge fluctuations, such that the spinon motion is twisted by the 
induced internal U(1) gauge flux. 
Similar ideas have been extended to the mixed valence compounds where 
the Fermi surface of neutral particles has been proposed~\cite{Senthil}, although
the thermal Hall measurement in SmB$_6$ or YbB$_{12}$ gives a zero result~\cite{Matsuda}. 
For strong Mott insulators, the charge gap 
is large and the charge fluctuation is strongly suppressed. This induction
of the internal U(1) gauge flux via strong charge fluctuations does not 
apply to the strong Mott regimes.

In the U(1) QSLs in the strong Mott regime, different physical mechanisms are needed
to understand the large thermal Hall effect. For the U(1) QSLs whose gauge flux is
related to the scalar spin chirality ${({\boldsymbol S}_i \times {\boldsymbol S}_j)
\cdot {\boldsymbol S}_k}$, we pointed out that, the combination of the Dzyaloshinskii-Moriya
interaction and a simple Zeeman coupling could generate an internal U(1) gauge flux,
and thus twist the motion of the spinons~\cite{Chen1901.01522,PhysRevResearch.1.013014}. 
This mechanism does not depend on the choices of the (bosonic) Schwinger spinons or the (fermionic) Abrikosov spinons. The (fermionic) Abrikosov spinons describe
more QSL states in 2D. The bosonic Schwinger spinon does not work for U(1) QSLs 
in 2D due to the confinement issue from the instanton effect. So for the Schwinger 
spinon description, this mechanism would only apply to the 3D U(1) QSL. In contrast, 
this mechanism broadly applies to the U(1) QSLs with the fermionic spinon description.

For the pyrochlore U(1) QSL that is also in the strong Mott regime, the relation
of the internal variable and the physical variable is much simpler than the one 
described in the previous paragraph. So the linear Zeeman coupling already induces 
an internal dual U(1) gauge flux and twist the motion of the ``magnetic monopoles.''

In general, for the QSLs with a {\sl continuous} gauge theory description, one key to 
resolve the mechanism for the thermal Hall effect is to understand the physical 
manifestation of the internal gauge flux and then the role of the external probes. 
This is related to the relation between the microscopic degrees of freedom and 
the emergent degrees of freedom in the lattice gauge theory formulation.

\subsection{Comparison with $\mathbb{Z}_2$ QSLs}


For $\mathbb{Z}_2$ QSLs, the above mechanism does not apply because the 
internal gauge flux is gapped and discrete and cannot be changed in a continuous
manner. An example would be the $\mathbb{Z}_2$ QSL from the Balents-Fisher-Girvin 
model~\cite{PhysRevB.65.224412}. Although the $\mathbb{Z}_2$ vison experiences a
dual background $\pi$ flux and the $S^z$-$S^z$ dynamical correlation has a spectral 
periodicity enhancement, a small magnetic field cannot modify this background flux 
continuously. Likewise, the spinons experience a background $0$ flux, and the 
magnetic field cannot change this flux continuously. Thus the mechanism in the 
previous subsection neither applies to the spinon nor to the vison. 
In $\mathbb{Z}_2$ QSLs, instead, it is the non-trivial band structure of 
matter field that directly contributes to the thermal Hall conductivity. 
A representative example would be the Kitaev model at the isotropic point 
where the spinons develop a gapless Dirac-type majorana ferminon band structure~\cite{KITAEV20062}. 
When the magnetic field is applied to the system, the field generates a mass
gap for the majorana fermions and creates a topological spinon band structure 
with a non-trivial Chern number. This is the origin of the thermal Hall effect 
for Kitaev QSL.

Another studied case~\cite{Subir1812.08792} are gapped $\mathbb{Z}_2$ QSLs 
with the Schwinger boson description.
The Dzyaloshinskii-Moriya interaction and the Zeeman coupling 
together breaks the time reversal symmetry and inversion symmetry. 
It was suggested that, using the Schwinger boson construction, 
the Dzyaloshinskii-Moriya interaction and the Zeeman coupling 
together generates a non-trivial Berry curvature distribution 
for the (gapped) bosonic spinon bands. At finite temperatures, the 
spinon bands are populated thermally, contributing to the thermal
Hall conductivity.

\subsection{Materials' survey}




The pyrochlore U(1) QSLs have been proposed for several rare-earth pyrochlore magnets.
Here we give a detailed discussion about the potential thermal Hall conductivity in 
some key representatives. 

We start with the non-Kramers doublets. Here the Tb family 
Tb$_2$Ti$_2$O$_7$~\cite{PhysRevLett.98.157204,PhysRevB.92.245114,PhysRevB.87.060408}
and the Pr family (Pr$_2$Zr$_2$O$_7$, Pr$_2$Sn$_2$O$_7$, 
Pr$_2$Hf$_2$O$_7$)~\cite{Kimura2012,Romain2017,PhysRevB.88.104421}
have been proposed as pyrochlore U(1) QSLs. The thermal Hall effect 
has been measured in Tb$_2$Ti$_2$O$_7$~\cite{Hirschberger106}, 
and inelastic neutron scattering measurement
has been performed on the Pr-based family~\cite{PhysRevLett.118.107206,PhysRevLett.118.087203,Sibille2018}. 
The continuous spectrum has been obtained experimentally. It was proposed that, 
the inelastic neutron scattering results for the non-Kramers doublets would contain 
the continuum of the ``magnetic monopoles'' from the duality arguments~\cite{PhysRevB.96.195127}. 
Other theory from the crystal field disorders of the non-Kramers doublets interpreted the excitation 
continuum differently~\cite{PhysRevLett.118.087203}. The
Tb$_2$Ti$_2$O$_7$ sample can become Ising ordered
once the stoichiometry of the sample is changed~\cite{PhysRevB.87.060408,PhysRevB.92.245114}.
Actually since Tb$^{3+}$ carries a non-Kramers doublet, the Ising order transition 
should be understood as the ``magnetic monopole'' condensation out of the 
U(1) QSL if the original disordered state is a U(1) QSL~\cite{PhysRevB.94.205107}. 
Therefore, both Tb-based and Pr-based rare-earth pyrochlore materials 
can be good candidates for the pyrochlore U(1) QSLs. 
We expect a non-trivial thermal Hall effect 
to be established in these candidate materials. 

The well-known Yb$_2$Ti$_2$O$_7$~\cite{PhysRevB.95.094422,
PhysRevLett.103.227202,PhysRevLett.119.057203,PhysRevLett.115.267208,
PhysRevLett.109.097205,PhysRevB.95.094407,
PhysRevB.96.214415,PhysRevB.92.064425,2019NatComm} 
is now under debate~\cite{PhysRevX.1.021002}.
Here the Yb$^{3+}$ ion is a Kramers ion and differs from the non-Kramers 
Pr$^{3+}$ ion. The actual low-temperature phase depends sensitively
on the preparation of the samples. For the physical point of view, 
it does not really matter strongly whether the magnetic ordered state 
of the system is proximate to the spin ice 
or not proximate to spin ice. The pyrochlore U(1) QSL can persist 
beyond the perturbative spin ice regime. A more sensible question 
would be whether Yb$_2$Ti$_2$O$_7$ is proximate to the pyrochlore 
U(1) QSL rather than proximate to the (perturbative) spin ice manifold. 
If the system is proximate to the pyrochlore U(1) QSL, then
TTHE of ``magnetic monopoles'' could be relevant and may even
persist to the weak ordered regime, despite the fact that 
the Zeeman coupling involves the transverse spin components. The Zeeman coupling
with the transverse spin components modifies the spinon dispersion and 
could provide a thermal Hall signal of spinons. As the spinons  
usually have much higher energy scales than the ``magnetic monopoles'',
we expect that the low-temperature thermal Hall effect is still dominated 
by the ``magnetic monopoles''.

Recently, Ce$_2$Zr$_2$O$_7$, Ce$_2$Sn$_2$O$_7$, and Ce$_2$Hf$_2$O$_7$
have been realized and proposed as QSLs~\cite{sibille1502candidate,dai1901experimental}. 
The Ce$^{3+}$ ion is also a Kramers ion of the dipole-octupole type~\cite{PhysRevLett.112.167203}, 
but differs from the Yb$^{3+}$ ion. Each state of the ground state 
doublet of the Ce$^{3+}$ ion is a one-dimensional irreducible 
representation of the D$_{3d}$ point group~\cite{PhysRevB.95.041106,Chen1902.07075}, 
while the two states of the Yb$^{3+}$ ion comprise a two-dimensional irreducible 
representation. It was suggested that, two distinct symmetry enriched U(1) QSLs, i.e.,
dipolar U(1) QSL and octupolar U(1) QSL, can be 
stabilized by studying the generic model for dipole-octupole doublets. 
The dipolar U(1) QSL is identical to the one obtained for the non-Kramers doublets
and the usual Kramers doublets. Since the external magnetic field primarily couples 
to the dipolar component at the linear level, if the dipolar U(1) QSL is stabilized,
then we expect the TTHE of ``magnetic monopoles.'' On the other hand, 
if the octupolar U(1) QSL is stabilized, 
the external magnetic field would modify the spinon band 
structure~\cite{PhysRevB.95.041106,Chen1902.07075} but 
would not change the dual U(1) flux for the ``magnetic monopoles,''
so we do not expect the TTHE for the ``magnetic monopoles.''

\section{Acknowledgments}

We acknowledge Xuefeng Sun from USTC for the discussion of experimental setups. 
XTZ acknowledges Ryuichi Shindou for previous collaborations on the
duality related topics. This work is supported by the Ministry of Science and 
Technology of China with Grant No.2016YFA0301001, 2016YFA0300500, 2018YFGH000095
and by General Research Fund (GRF) No.17303819 from the Research Grants Council 
of Hong Kong.

\appendix 

\section{Duality transformation} 
\label{AppSec1}

Start from the ring exchange Hamiltonian in Eq.~\eqref{ring}, 
we rewrite in terms of a particle number $n_i$ 
(integer-valued) and a conjugated phase $\phi_i$ ,
\begin{equation}
\begin{aligned}
& \tau^{\pm}_i= e^{\pm i\phi_i} \\
& \tau^z_i = n_i - \frac{1}{2}
\end{aligned}
\end{equation}
which satisfy the commutation relation 
\begin{equation}
[\phi_i,n_i]= i
\label{comm_1}
\end{equation}

Moreover, $\tau^z_i$ takes the eigenvalue of $\pm 1/2$. 
To ensure the Hilbert space is not enlarged, we add a 
constrain term $(n_i-\frac{1}{2})^2$ with a strength $U$.
The particle number takes values $n_i=0,1$, and, 
we obtain a Hamiltonian 
\begin{equation}
\begin{aligned}
{\cal H}_{\rm ring}
= & -K  \sum_{\hexagon_p} \cos\big( \phi_1-\phi_2 + \phi_3 - \phi_4 + \phi_5 - \phi_6 \big) \\
& + \frac{U}{2}\sum_{i} (n_i-\frac{1}{2})^2 \\
\end{aligned}
\label{phi_n}
\end{equation}
Now, we transform to the electric field and gauge field, 
which are defined on the diamond lattice, (see Fig.~\ref{2})
\begin{equation}
\begin{aligned}
& A_{rr^\prime}= \epsilon_r \phi_{rr^\prime} \\
& E_{rr^\prime}= \epsilon_r n_{rr^\prime}  \\
\end{aligned}
\label{A_E_phi_n}
\end{equation}
where the pyrochlore site $i$ sits in the middle of the link $rr^\prime$. 
$\epsilon_r=+1(-1)$ for diamond lattice type-${\rm \RNum{1}}({\rm \RNum{2}})$.
Thus, the variables are anti-symmetric ${\cal G}_{r^\prime r}=- {\cal G}_{rr^\prime}, \ \ {\cal G}=A,E$.
And, the commutation relation follows from Eq.~\eqref{comm_1}
\begin{equation}
[A_{rr^\prime},E_{rr^\prime}] = \epsilon_r^2 [\phi_i, n_i] = i
\label{A_E}
\end{equation}
We fix the branch-cut for the $2\pi$-periodic variable as $A_{rr^\prime} \in [-\pi,+\pi)$,
so that a lattice curl of this variable remains non-vanishing,
\begin{equation}
{\rm curl} A(\textsf{r} \textsf{r}^\prime)= \sum_{rr^\prime \in \hexagon_{d(\textsf{r} \textsf{r}^\prime)}} A_{rr^\prime}
\end{equation}
where the original diamond hexagon $\hexagon_{d(\textsf{r} \textsf{r}^\prime)}$ 
is labelled by the dual diamond link $\textsf{r} \textsf{r}^\prime$ that penetrates the hexagon.
The phase terms in Eq.~\eqref{phi_n} is expressed in an elegant way,
\begin{equation}
\begin{aligned}
{\cal H}_{\rm LGT}[A,E]
= &- K  \sum_{\hexagon_{d(\textsf{r} \textsf{r}^\prime)}} \cos( {\rm curl} A_{\textsf{r} \textsf{r}^\prime}) \\
& + \frac{U}{2}\sum_{rr^\prime} \big(E_{rr^\prime}-\frac{\epsilon_r}{2} \big)^2 \\
\end{aligned}
\end{equation}
which has been presented in Eq.~\eqref{LGT} as a lattice gauge theory.
And, the corresponding action reads,
\begin{equation}
\begin{aligned}
{\cal S}_{\rm LGT}[A,E]
= & \sum_{rr^\prime} A_{rr^\prime} \partial_t E_{rr^\prime} + {\cal H}_{\rm LGT}[A,E]\\
\end{aligned}
\label{action_LGT}
\end{equation}
Along this line of derivation, we should keep track of 
a ``2-in-2-out'' configuration of the spins in a pyrochlore tetrahedra,
\begin{equation}
\sum_{i\in {\rm teh}_r} \tau^z_i =0
\end{equation}
where pyrochlore sites $i$ belong to the tetrahedra labelled by its center $r$-site.
As a result, the electric field is imposed with a constraint
\begin{equation}
\begin{aligned}
 {\rm div}E(r)\equiv \sum_{r^\prime\in r} E_{rr^\prime} =\sum_{i\in r} n_i = 2\epsilon_r
\end{aligned}
\label{div_E}
\end{equation}
where the summation defines a lattice divergence,
and $r^\prime\in r$ refers to the four nearest neighbor 
original site of a given original site $r$.

Next, we transform the LGT to a dual theory by defining
\begin{equation}
\begin{aligned}
& {\rm curl} a_{rr^\prime}  = E_{rr^\prime} - E^0_{rr^\prime} \\
& B_{\textsf{r} \textsf{r}^\prime} = {\rm curl} A_{\textsf{r} \textsf{r}^\prime} \\
\end{aligned}
\label{dual_variables}
\end{equation}
where a magnetic field $a_{\textsf{r} \textsf{r}^\prime}$ (integer-valued) 
and dual gauge field $B_{\textsf{r} \textsf{r}^\prime}$ ($2\pi$-periodic).
The curl of the dual gauge field is related to the electric field, therefore the z-component spin.
It is valid for a gauge invariant quantity to represent a physical one.
Since the dual gauge field is integer-valued, 
we expect no divergence for its lattice curve.
On the other hand, as dictated in Eq.~\eqref{div_E}, 
the electric field has a non-vanishing divergence.
A background electric field is introduced to ensure the divergencelessness of the dual gauge field.
We pick a particular configuration within the 2-in-2-out spin ice manifold,
\begin{equation}
\begin{aligned}
E^0_{r,r+\epsilon_r e_0} = & E^0_{r,r+\epsilon_r e_1} =\epsilon_r\\ 
E^0_{r,r+\epsilon_r e_2} = & E^0_{r,r+\epsilon_r e_3} =0 \\ 
\end{aligned}
\label{E_0}
\end{equation}
The dual lattice gauge theory is written as,
\begin{equation}
\begin{aligned}
{\cal H}_{\rm dual}[a,B]=
& \sum_{rr^\prime} \frac{U}{2} ({\rm curl}\ a_{rr^\prime}-\bar{E}_{rr^\prime})^2 - K \sum_{ \textsf{r} \textsf{r}^\prime}  \cos B_{\textsf{r} \textsf{r}^\prime} \\
\end{aligned}
\end{equation}
where the background electric field is defined in Eq.~\eqref{E_bar}.
The corresponding action follows from Eq.~\eqref{action_LGT}
\begin{equation}
\begin{aligned}
{\cal S}_{\rm dual}[a,B]
= &\sum_{ rr^\prime} A_{rr^\prime}\partial_t ({\rm curl}\ a_{rr^\prime}+{E}^0_{rr^\prime}) +{\cal H}_{\rm dual}[a,B]\\
= &\sum_{\textsf{r} \textsf{r}^\prime} B_{ \textsf{r} \textsf{r}^\prime} \partial_t (a_{ \textsf{r} \textsf{r}^\prime} + a^0_{ \textsf{r} \textsf{r}^\prime}) +{\cal H}_{\rm dual}[a,B]\\
\end{aligned}
\end{equation}
where $a^0_{ \textsf{r} \textsf{r}^\prime}$ is the vector 
potential responsible for the electric field ${E}^0_{rr^\prime}= {\rm curl} a^0_{rr^\prime}$.
And, in the second equality we have exchanged the sequence of 
the summation over original and dual lattices,
\begin{equation}
\sum_{ rr^\prime} \sum_{\textsf{r} \textsf{r}^\prime \in \hexagon^\ast_{d(rr^\prime)}} = \sum_{\textsf{r} \textsf{r}^\prime} \sum_{ rr^\prime  \in \hexagon_{d(\textsf{r} \textsf{r}^\prime)}} 
\end{equation}

The divergence of the magnetic field is non-zero by definition
\begin{equation}
\begin{aligned}
{\rm div} B_{\textsf{r}} 
\equiv & {\rm div} \cdot {\rm curl} A(\textsf{r}) \\
= & \sum_{\textsf{r}^\prime \in \textsf{r}}\sum_{rr^\prime \in \hexagon_{d(\textsf{r}\textsf{r}^\prime)}} {A_{rr^\prime}}_{[-\pi,\pi)} \\
= & 2\pi {\cal Z} \\
\end{aligned}
\label{div_B}
\end{equation}
The ``magnetic monopole'' number operator is defined as 
the topological defect of this magnetic field,
\begin{equation}
\begin{aligned}
N_{\textsf{r}} \equiv \frac{1}{2\pi}{\rm div} B_{\textsf{r}} \\
\end{aligned}
\label{N}
\end{equation}
which takes integer values. The commutation relation between the dual variables 
can be derived from Eq.~\eqref{A_E}, 
\begin{equation}
\big[ \sum_{rr^\prime \in \hexagon_{d(\textsf{r}\textsf{r}^\prime)}} 
A_{rr^\prime}, E_{rr^\prime} \big] = i 
\end{equation}
so that, we have
\begin{equation}
\begin{aligned}
& \big[ B_{\textsf{r}\textsf{r}^\prime}, 
\sum_{\textsf{r}_1\textsf{r}^\prime_1 \in \hexagon^\ast_{d(rr^\prime)}} 
a_{\textsf{r}_1 \textsf{r}_1^\prime} \big] = i, \ \ \ 
\textsf{r}\textsf{r}^\prime\in \hexagon^\ast_{d(rr^\prime)} \\
\end{aligned}
\label{a_B_1}
\end{equation}
Particularly, we can make a convenient choice
\begin{equation}
\begin{aligned}
\big[ B_{\textsf{r}\textsf{r}^\prime},a_{\textsf{r}_1\textsf{r}_1^\prime} \big] =
\left\{\begin{array}{cc}  
i, &  \ \ \ \ \ \textsf{r}_1\textsf{r}_1^\prime=\textsf{r}\textsf{r}^\prime \\
0, & \ \ \ \ \ \textsf{r}_1\textsf{r}_1^\prime \ne \textsf{r}\textsf{r}^\prime \\
\end{array}\right.  
\end{aligned}
\label{a_B_2}
\end{equation}
which we have used in and below Eq.~\eqref{a_B}.
Finally, we note that the two dual variables are anti-symmetric
with respect to exchanging the lattice sites.
This fact follows from the definitions in Eq.~\eqref{A_E_phi_n} 
and Eq.~\eqref{dual_variables}. 

From Eq.(\ref{div_B}) and Eq.(\ref{N}), we see that the ``magnetic monopole'' is the
topological defect of the dual vector gauge potential in the
compact U(1) quantum electrodynamics and has no classical
analog. Even though the spinon and the ``magnetic monopole''
can be interchanged by the electromagnetic duality of the 
lattice gauge theory, the ``magnetic monopole'' might be more
close in spirit to Dirac's magnetic monopole  from the
original definition and theory of the pyrochlore U(1) QSL\cite{PhysRevB.69.064404}.

So far, we have derived the dual gauge theory.
The commutation relation of variables is properly kept along the way.
And, we have identified the ``magnetic monopole'' number operator,
however, a conjugated phase operator of the ``magnetic monopole'' 
is missing in the present formulation.
Moreover, the dual gauge field is a discretized variable, 
which is cumbersome to deal with in terms of standard field theory methods.
Fortunately, during the process of ``softening'' the dual gauge field,
we can introduce the phase operator of the ``magnetic monopole'' in a natural way.
And, we are able to establish a commutation relation between the introduced phase variable
and the ``magnetic monopole'' number operator.

\section{``Variable-soften'' procedure} 
\label{AppSec2}

The model describes a confinement-deconfinement phase transition
due to the discreteness of the dual U(1) gauge field.
Otherwise, the partition function is basically a trivial Gaussian model.
Let us consider the dual gauge field part of the partition function,
\begin{equation}
\begin{aligned}
Z[a]\equiv & \sum_{\{ a_{\textsf{r} \textsf{r}^\prime}\}} e^{-\sum_{rr^\prime} \frac{U}{2}  ( {\rm curl}\ a_{rr^\prime} - \bar{E}_{rr^\prime})^2} \\
= & \int {\mathcal D}{a} \sum_{\{ { p}_{rr^\prime} \}} \   e^{-\sum_{rr^\prime} \frac{U}{2}  ( {\rm curl}\ a_{rr^\prime} - \bar{E}_{rr^\prime})^2} \\
\ \ \ \ \ \ \ \ \ \ \ \ \ \ \ \ \ \ \ & \times e^{i2\pi \sum_{rr^\prime} {\rm curl}\ a_{rr^\prime} \cdot p_{rr^\prime}  }  \\
\end{aligned}
\end{equation}
where we have the used the Poisson's resummation rule to leverage the discreteness of $ a_{\textsf{r} \textsf{r}^\prime} $,
\begin{equation}
\sum_{m=-\infty}^{+\infty} e^{i2\pi m x}= \sum_{n=-\infty}^{+\infty} \delta(x-n)
\end{equation}
We can further transform the expression
\begin{equation}
\begin{aligned}
Z[a]
= & \int {\mathcal D}{a} \sum_{\{ { p}_{rr^\prime} \}} \   e^{-\sum_{rr^\prime} \frac{U}{2}  ( {\rm curl}\ a_{rr^\prime} - \bar{E}_{rr^\prime})^2} \\
& \times e^{i2\pi  \sum_{\textsf{r} \textsf{r}^\prime} a_{\textsf{r} \textsf{r}^\prime} \cdot {\rm curl}\ p_{\textsf{r} \textsf{r}^\prime}  }  \\
\end{aligned}
\end{equation}
 by manipulating the two summations involved,
\begin{equation}
\begin{aligned}
\sum_{rr^\prime} {\rm curl }a_{rr^\prime} \cdot p_{rr^\prime}
= & \sum_{rr^\prime}\sum_{\textsf{r} \textsf{r}^\prime \in \hexagon^\ast_{d(rr^\prime)}} a_{\textsf{r} \textsf{r}^\prime} \cdot p_{rr^\prime} \\
= & \sum_{\textsf{r} \textsf{r}^\prime}\sum_{ rr^\prime \in \hexagon_{d(\textsf{r} \textsf{r}^\prime)}} a_{\textsf{r} \textsf{r}^\prime} \cdot p_{rr^\prime} \\
= & \sum_{\textsf{r} \textsf{r}^\prime} a_{\textsf{r} \textsf{r}^\prime} \cdot {\rm curl}\ p_{\textsf{r} \textsf{r}^\prime} \\
\end{aligned}
\label{a_p}
\end{equation}
Importantly, the dual gauge field is anti-symmetric, 
i.e., $a_{\textsf{r}^\prime \textsf{r}}=-a_{\textsf{r} \textsf{r}^\prime}$.
The curl of the auxiliary field ${\rm curl}\ p_{\textsf{r} \textsf{r}^\prime}$ is anti-symmetric as well,
so the summation in Eq.~\eqref{a_p} gives non-vanishing result.
Moreover, this curl is divergentless, since $p_{rr^\prime}$ is an integer-valued variable.
The divergentless and anti-symmetric properties can be made explicit in the path integral formulation,
\begin{equation}
\begin{aligned}
Z[a]
= & \int {\mathcal D}{a} e^{-\sum_{rr^\prime} \frac{U}{2}  ( {\rm curl}\ a_{rr^\prime} - \bar{E}_{rr^\prime})^2 } \times \{ ...\}\\
 \{ ...\}= & \sum_{\{ {M}^{\rm asym}_{\textsf{r}  \textsf{r}^\prime} \}} \delta\Big[ {\rm div} M^{\rm asym}(\textsf{r})\Big]\   e^{i2\pi  \sum_{\textsf{r} \textsf{r}^\prime} a_{\textsf{r} \textsf{r}^\prime} \cdot M^{\rm asym}_{\textsf{r} \textsf{r}^\prime}  }  \\
\end{aligned}
\end{equation}
where $M^{\rm asym}_{\textsf{r} \textsf{r}^\prime}$ is anti-symmetric, integer-valued variable, 
and the lattice divergence is ${\rm div} M^{\rm asym}(\textsf{r}) = \sum_{ \textsf{r}^\prime \in  \textsf{r}}M^{\rm asym}_{\textsf{r} \textsf{r}^\prime}$. 
The delta function can be removed by introducing another auxiliary field $\theta_{\textsf{r}}$,
\begin{equation}
\begin{aligned}
 \{ ...\}= & \sum_{\{ {M}^{\rm asym}_{\textsf{r}  \textsf{r}^\prime} \}}  e^{i2\pi  \sum_{\textsf{r} \textsf{r}^\prime} a_{\textsf{r} \textsf{r}^\prime} \cdot M^{\rm asym}_{\textsf{r} \textsf{r}^\prime}  } 
 \int {\cal D}\theta \ e^{i \sum_{\textsf{r}} {\rm div} M^{\rm asym}_{\textsf{r}} \cdot \theta_{\textsf{r}} } \\
\end{aligned}
\end{equation}
Now, we are in the position to remove the anti-symmetric condition
\begin{equation}
\begin{aligned}
 \sum_{\textsf{r}} {\rm div} M^{\rm asym}_{\textsf{r}} \cdot \theta_{\textsf{r}} 
= & \sum_{\textsf{r}} \sum_{\textsf{r}^\prime \in \textsf{r}} M^{\rm asym}_{\textsf{r}\textsf{r}^\prime} \cdot \theta_{\textsf{r}} \\
= & \sum_{\textsf{r}\textsf{r}^\prime}  (M_{\textsf{r} \textsf{r}^\prime} - M_{\textsf{r}^\prime \textsf{r}}) \theta_{\textsf{r}} \\
= &  \sum_{\textsf{r}\textsf{r}^\prime}  M_{\textsf{r} \textsf{r}^\prime} ( \theta_{\textsf{r}}- \theta_{\textsf{r}^\prime}) \\
\end{aligned}
\label{M_theta}
\end{equation}
So that, we arrive at an elegant expression, which is similar to the result in literature~\cite{herbut}
\begin{equation}
\begin{aligned}
 \{ ...\}= &\sum_{\{ {M}_{\textsf{r}  \textsf{r}^\prime} \}}    
e^{i2\pi  \sum_{\textsf{r} \textsf{r}^\prime} a_{\textsf{r} \textsf{r}^\prime} \cdot M_{\textsf{r} \textsf{r}^\prime}  } 
e^{i  \sum_{\textsf{r}\textsf{r}^\prime}  M_{\textsf{r} \textsf{r}^\prime} ( \theta_{\textsf{r}}- \theta_{\textsf{r}^\prime})} \\
\end{aligned}
\end{equation}
Following the series of transformation and perform a Villain approximation, 
we end up with the dual theory in Eq.~\eqref{dual_ham}. The 
\begin{equation}
\begin{aligned}
{\cal H}_{\rm dual}[\theta, a,B] = & \sum_{r r^\prime} \frac{U}{2}  ( {\rm curl}\ a_{rr^\prime} - \bar{E}_{rr^\prime})^2 
- \sum_{ \textsf{r} \textsf{r}^\prime} K \cos B_{\textsf{r} \textsf{r}^\prime}  \\
& - t \sum_{\textsf{r} \textsf{r}^\prime} \cos(\theta_{\textsf{r}} - \theta_{\textsf{r}^\prime}+ 2\pi a_{\textsf{r} \textsf{r}^\prime} )  \\
{\rm cond}: \ \ \ & {\rm div} B(\textsf{r}) = 2\pi N_{\textsf{r}} \\
\end{aligned}
\label{dual_Ham_cond}
\end{equation}
where a parameter $t$ is added as a chemical potential term for the $M_{\textsf{r} \textsf{r}^\prime}$.
So far, we have resolve the discreteness issue of the dual gauge field
by introducing a phase field $\theta_{\textsf{r}}$.
At the moment, the physical meaning of this variable is not clear,
namely, the commutation relation with the other variables are not given.

Further progress is made by manipulating the condition in Eq.~\eqref{dual_Ham_cond}.
The full partition function and action are given by,
\begin{equation}
\begin{aligned}
Z= & \int {\mathcal D}\theta {\mathcal D}a \int_{\rm cond}{\mathcal D} B e^{i \sum_{\textsf{r}\textsf{r}^\prime}B_{\textsf{r}\textsf{r}^\prime} \partial_\tau (a_{\textsf{r}\textsf{r}^\prime}+a^0_{\textsf{r}\textsf{r}^\prime})  - {\cal H}_{\rm dual}[\theta, a, B]} \\
\equiv & \int {\mathcal D}\theta {\mathcal D}a \int_{\rm cond}{\mathcal D} B e^{ -{\cal S}_{\rm dual}[\theta, a, B] }  \\
\end{aligned}
\label{full_partition}
\end{equation}
where the condition in the integral can be made explicit by inserting another delta function,
\begin{equation}
\begin{aligned}
Z= & \int {\mathcal D}\theta {\mathcal D}a \int{\mathcal D} B \delta\big[ {\rm div} B(\textsf{r}) - 2\pi N_{\textsf{r}}\big] e^{ -{\cal S}_{\rm dual}[\theta, a, B] } \\
= & \int {\mathcal D}\theta {\mathcal D}a {\mathcal D} B {\mathcal D} \Lambda \  e^{i \sum_{\textsf{r}} \Lambda_{\textsf{r}} ({\rm div} B_{\textsf{r}} - 2\pi N_{\textsf{r}})} e^{ -{\cal S}_{\rm dual}[\theta, a, B] }  \\
\end{aligned}
\label{gauge_generator}
\end{equation}
where $G(\Lambda)=e^{i \sum_{\textsf{r}} \Lambda_{\textsf{r}} ({\rm div} B_{\textsf{r}} - 2\pi N_{\textsf{r}})}$ is regarded as a gauge fixing generator~\cite{fisher}, 
which can be transformed in the similar way as in Eq.~\eqref{M_theta}, 
\begin{equation}
G(\Lambda)=e^{i \sum_{\textsf{r} \textsf{r}^\prime} B_{\textsf{r}\textsf{r}^\prime}(\Lambda_{\textsf{r}} - \Lambda_{\textsf{r}^\prime} ) } e^{- i2\pi \sum_{\textsf{r}} N_{\textsf{r}} \Lambda_{\textsf{r}}} 
\label{G_func}
\end{equation}
This function generates a gauge transformation for functions 
involving the dual gauge field and the phase variable,
\begin{equation}
\begin{aligned}
& G(\Lambda) {\cal S}_{\rm dual}( a_{\textsf{r}\textsf{r}^\prime},\theta_{\textsf{r}} , B_{\textsf{r}\textsf{r}^\prime}) G^\dagger(\Lambda) \\
= & {\cal S}_{\rm dual}\big[ a_{\textsf{r}\textsf{r}^\prime}+ ( \Lambda_{\textsf{r}}- \Lambda_{\textsf{r}^\prime}),\theta_{\textsf{r}} + 2\pi \Lambda_{\textsf{r}}, B_{\textsf{r}\textsf{r}^\prime} \big]\\
\end{aligned}
\label{gauge_trans}
\end{equation}
under the condition that the following commutation relation is satisfied,
\begin{equation}
[\theta_{\textsf{r}}, N_{\textsf{r}^\prime}] = i \delta_{\textsf{r},\textsf{r}^\prime}
\label{N_theta}
\end{equation}
The transformed action is equivalent to the original one by absorbing the field $\Lambda_{\textsf{r}}$, we have
\begin{equation}
\begin{aligned}
Z= & \int {\mathcal D}\theta {\mathcal D}a {\mathcal D} B {\mathcal D} \Lambda \ e^{ -{\cal S}_{\rm dual}[\theta, a, B] }e^{i \sum_{\textsf{r}} \Lambda_{\textsf{r}} ({\rm div} B_{\textsf{r}} - 2\pi N_{\textsf{r}})}  \\
= & \int {\mathcal D}\theta {\mathcal D}a \int{\mathcal D} B  e^{ -{\cal S}_{\rm dual}[\theta, a, B] } \delta\big[ {\rm div} B(\textsf{r}) - 2\pi N_{\textsf{r}}\big]\\
= & \int {\mathcal D}\theta {\mathcal D}a \int_{\rm cond}{\mathcal D} B e^{ -{\cal S}_{\rm dual}[\theta, a, B] }  \\
\end{aligned}
\end{equation}
where the action is intact after applying the gauge generator obeying the commutation rule in Eq.~\eqref{N_theta}.
Therefore, the variable $\theta$ admits a physical meaning of the conjugated phase of the magnetic monopole.
$e^{i\theta}(e^{-i\theta})$ is the creation(annihilation) operator for the magnetic monopole.

Conclusively, we finish the task of softening the dual gauge field in the dual theory, meanwhile,
introducing the magnetic monopole phase variable. 
We emphasize on the peculiar definition of curl and divergence in the diamond lattice structure,
and the anti-symmetric property of the link variables.

\section{Thermal Hall Current Operator} 
\label{AppSec3}

In Sec.~\ref{sec3}, we present the result for the thermal Hall current at the mean-field level.
Here, we derive a compact expression for the thermal Hall current in the presence of gauge fluctuation.
We start from the same energy continuity equation as in Eq.~\eqref{continuity}, 
yet with a different energy density operator,
\begin{equation}
\begin{aligned}
{\cal H}_{\rm dual} = &   \sum_{\textsf{r}} {\cal H}_{\textsf{r}} \\
 {\cal H}_{\textsf{r}}= &   \sum_{\textsf{r}^\prime \in \textsf{r} }  \Big\{ \frac{U}{2} \sum_{rr^\prime \in \hexagon_{d(\textsf{r}\textsf{r}^\prime)}}  ( {\rm curl}\ a_{rr^\prime} - \bar{E}_{rr^\prime})^2 - \frac{K}{2}  B_{\textsf{r} \textsf{r}^\prime}^2 \\
& - \frac{t}{2} e^{i (\theta_{\textsf{r}} - \theta_{\textsf{r}^\prime}+ 2\pi a_{\textsf{r} \textsf{r}^\prime} )} + {\rm H.c.} \Big\} \\
\end{aligned}
\end{equation}
where we have kept the rotor variable $e^{i\theta_{\textsf{r}}}$ 
instead of the boson field used in the main text. 
And, the summation of the first term comes from,
\begin{equation}
\begin{aligned}
\sum_{rr^\prime} 
\simeq \sum_{\textsf{r}\textsf{r}^\prime}  \sum_{rr^\prime \in \hexagon_{d(\textsf{r}\textsf{r}^\prime)}}
\simeq  \sum_{\textsf{r}} \sum_{\textsf{r}^\prime \in \textsf{r}}  \sum_{rr^\prime \in \hexagon_{d(\textsf{r}\textsf{r}^\prime)}} 
\end{aligned}
\end{equation}

Next, we evaluate the time partial derivative of the energy density
\begin{equation}
\begin{aligned}
& \dot{\cal H}_{\textsf{r}}= -i \big[{\cal H}_{\textsf{r}}, {\cal H}_{\rm dual} \big] \\
= &  i\frac{UK}{4} \sum_{\textsf{r}^\prime \in \textsf{r} } \sum_{r r^{\prime} \in \hexagon_{d(\textsf{r}\textsf{r}^\prime)}}  \sum_{\textsf{r}_1\textsf{r}^\prime_1  }  \big[ ( {\rm curl}\ a_{rr^\prime} - \bar{E}_{rr^\prime})^2 ,   B^2_{\textsf{r}_1 \textsf{r}^\prime_1}  \big] \\
& +\sum_{\textsf{r}^\prime \in \textsf{r} }   \sum_{rr^\prime } \big[  B^2_{\textsf{r} \textsf{r}^\prime}  , ( {\rm curl}\ a_{rr^\prime} - \bar{E}_{rr^\prime})^2 \big] \\
\end{aligned}
\end{equation}
The 1st and 2nd commutators are calculated respectively. 
In the 1st term, we use the commutation relation in Eq.~\eqref{a_B_1},
while in the 2nd term, we use a modified version of commutation relation,  i.e., $ \big[ B_{\textsf{r}\textsf{r}^\prime}, {\rm curl}\ a_{rr^\prime} \big] 
= i,\ rr^\prime \in \hexagon_{d(\textsf{r}\textsf{r}^\prime)} $.

\begin{widetext}
\begin{equation}
\begin{aligned}
 \{ {\rm 1st} \}\equiv & \sum_{rr^\prime \in  \hexagon_{d(\textsf{r}\textsf{r}^\prime)}} \sum_{\textsf{r}_1\textsf{r}^\prime_1  } \big[ ( {\rm curl}\ a_{rr^\prime} - \bar{E}_{rr^\prime})^2 ,   B^2_{\textsf{r}_1 \textsf{r}^\prime_1}  \big] \\
= &     \sum_{rr^\prime  \in \hexagon_{d(\textsf{r}\textsf{r}^\prime)}} \sum_{\textsf{r}_1\textsf{r}^\prime_1  } ( {\rm curl}\ a_{rr^\prime} - \bar{E}_{rr^\prime}) \big[ {\rm curl}\ a_{rr^\prime},   B^2_{\textsf{r}_1 \textsf{r}^\prime_1}  \big]  +\big[ {\rm curl}\ a_{rr^\prime},   B^2_{\textsf{r}_1 \textsf{r}^\prime_1}  \big] ( {\rm curl}\ a_{rr^\prime} - \bar{E}_{rr^\prime}) \\
= &  \sum_{rr^\prime  \in \hexagon_{d(\textsf{r}\textsf{r}^\prime)}} \sum_{\textsf{r}_1\textsf{r}_1^\prime} \delta[\textsf{r}_1\textsf{r}_1^\prime \in \hexagon^\ast_{d(rr^\prime)}] \Big\{ ( {\rm curl}\ a_{rr^\prime} - \bar{E}_{rr^\prime}) (-2i B_{\textsf{r}_1 \textsf{r}^\prime_1}) + (-2i B_{\textsf{r}_1 \textsf{r}^\prime_1}) ( {\rm curl}\ a_{rr^\prime} - \bar{E}_{rr^\prime}) \Big\} \\
= &  \sum_{rr^\prime  \in \hexagon_{d(\textsf{r}\textsf{r}^\prime)}} \sum_{\textsf{r}_1\textsf{r}_1^\prime \in \hexagon^\ast_{d(rr^\prime)}}   ( {\rm curl}\ a_{rr^\prime} - \bar{E}_{rr^\prime}) (-2i B_{\textsf{r}_1 \textsf{r}^\prime_1}) + (-2i B_{\textsf{r}_1 \textsf{r}^\prime_1}) ( {\rm curl}\ a_{rr^\prime} - \bar{E}_{rr^\prime}) \\
= &-2i  \sum_{rr^\prime  \in \hexagon_{d(\textsf{r}\textsf{r}^\prime)}}    ( {\rm curl}\ a_{rr^\prime} - \bar{E}_{rr^\prime}) {\rm curl}\ B_{rr^\prime}  +  {\rm curl}\ B_{rr^\prime} ( {\rm curl}\ a_{rr^\prime} - \bar{E}_{rr^\prime}) \\
\end{aligned}
\end{equation}
\end{widetext}

\begin{widetext}
\begin{equation}
\begin{aligned}
 \{ {\rm 2nd} \}\equiv &  \sum_{\textsf{r}_1\textsf{r}_1^\prime} \sum_{rr^\prime  \in  \hexagon_{d(\textsf{r}_1\textsf{r}_1^\prime)}} \big[  B^2_{\textsf{r} \textsf{r}^\prime}  , ( {\rm curl}\ a_{rr^\prime} - \bar{E}_{rr^\prime})^2 \big] \\
= & \sum_{\textsf{r}_1\textsf{r}_1^\prime} \sum_{rr^\prime  \in \hexagon_{d(\textsf{r}_1\textsf{r}_1^\prime)}}\big[  B^2_{\textsf{r} \textsf{r}^\prime}  ,{\rm curl}\ a_{rr^\prime}  \big]  ( {\rm curl}\ a_{rr^\prime} - \bar{E}_{rr^\prime}) + ( {\rm curl}\ a_{rr^\prime} - \bar{E}_{rr^\prime}) \big[  B^2_{\textsf{r} \textsf{r}^\prime}  ,{\rm curl}\ a_{rr^\prime}  \big] \\
= &  \sum_{\textsf{r}_1\textsf{r}_1^\prime} \sum_{rr^\prime  \in \hexagon_{d(\textsf{r}_1\textsf{r}_1^\prime)}}  \delta[rr^\prime \in \hexagon_{d(\textsf{r} \textsf{r}^\prime)}] \Big\{ (2i  B_{\textsf{r} \textsf{r}^\prime}) ( {\rm curl}\ a_{rr^\prime} - \bar{E}_{rr^\prime})  +  ( {\rm curl}\ a_{rr^\prime} - \bar{E}_{rr^\prime}) (2i B_{\textsf{r} \textsf{r}^\prime}) \Big\} \\
= &  \sum_{\textsf{r}_1\textsf{r}_1^\prime} \sum_{rr^\prime} \delta[rr^\prime  \in \hexagon_{d(\textsf{r}_1\textsf{r}_1^\prime)}]  \delta[rr^\prime \in \hexagon_{d(\textsf{r} \textsf{r}^\prime)}] \Big\{ (2i  B_{\textsf{r} \textsf{r}^\prime}) ( {\rm curl}\ a_{rr^\prime} - \bar{E}_{rr^\prime})  +  ( {\rm curl}\ a_{rr^\prime} - \bar{E}_{rr^\prime}) (2i B_{\textsf{r} \textsf{r}^\prime}) \Big\} \\
= &   \sum_{rr^\prime} \delta[rr^\prime \in \hexagon_{d(\textsf{r} \textsf{r}^\prime)}]  \sum_{\textsf{r}_1\textsf{r}_1^\prime} \delta[ \textsf{r}_1\textsf{r}_1^\prime \in \hexagon^\ast_{d(rr^\prime )}]  \Big\{ ...... \Big\} \\
= &   \sum_{rr^\prime \in \hexagon_{d(\textsf{r} \textsf{r}^\prime)}}  \sum_{\textsf{r}_1\textsf{r}_1^\prime \in \hexagon^\ast_{d(rr^\prime )}}   \Big\{ ...... \Big\} \\
= & 2i   \sum_{rr^\prime \in \hexagon_{d(\textsf{r}\textsf{r}^\prime)}}   \sum_{\textsf{r}_1\textsf{r}_1^\prime \in \hexagon^\ast_{d(rr^\prime )}}  B_{\textsf{r} \textsf{r}^\prime} ( {\rm curl}\ a_{rr^\prime} - \bar{E}_{rr^\prime})  +  ( {\rm curl}\ a_{rr^\prime} - \bar{E}_{rr^\prime})  \sum_{\textsf{r}_1\textsf{r}_1^\prime \in \hexagon^\ast_{d(rr^\prime )}}  B_{\textsf{r} \textsf{r}^\prime} \\
\end{aligned}
\end{equation}
\end{widetext}

Collecting terms from the two terms, we end up with an expression 
which involves the gauge fields in addition to the contribution from 
the matter field (``magnetic monopole'').
And, there is no matter-gauge coupling in the expression of the current, 
since the two sets of variables commute with each other.
Combining the mean-field solution in Eq.~\eqref{J_0} and 
this gauge field solution, we arrive at a total thermal Hall current operator,
\begin{equation}
{\cal J}^{0,E}_{\rm tot}= {\cal J}^{0,E}+ \delta {\cal J}^{0,E}
\end{equation}

\section{Basis convention} 
\label{AppSec4}

The thermal Hall effect considered in the main text takes place in the ($x,y$)-plane.
Here, we define a cuboid brillouin zone in this absolute coordinate.\\

The basis vectors of the diamond links are
\begin{equation}
\begin{aligned}
& {e}_0 = \frac{1}{\sqrt{3}} (+1,+1,+1);\ \  {e}_1 = \frac{1}{\sqrt{3}} (+1,-1,-1) \\
& {e}_2 = \frac{1}{\sqrt{3}} (-1,+1,-1); \ \ {e}_3 = \frac{1}{\sqrt{3}} (-1,-1,+1) \\
\end{aligned}
\label{emu}
\end{equation}

Within the same set of coordinate, 
we write down the real-space basis vectors $a_\nu\ (\nu=1,2,3)$ 
of the 3D super-lattice 
\begin{equation}
\begin{aligned}
& a_1= 4(e_0-e_3) = \frac{8}{\sqrt{3}} (1,1,0), \\
& a_2= e_0-e_1 = \frac{2}{\sqrt{3}} (0,1,1), \\
& a_3= e_3- e_2 = \frac{2}{\sqrt{3}} (0,0,1).\\
\end{aligned}
\label{a_nu}
\end{equation}

The reciprocal Wigner-Seitz brillouin zone (BZ) is spanned by $b_\nu\ (\nu=1,2,3)$,
\begin{equation}
\begin{aligned}
b_1= & \frac{\sqrt{3} \pi}{4}(1,0,0),\\
b_2= & \sqrt{3}\pi (-1,1,0),\\ 
b_3= & \sqrt{3}\pi  (1,-1,1)\\
\end{aligned}
\label{b_nu}
\end{equation}

The shape of the BZ can be adjusted to a cuboid one 
covering the same amount of volume in the momentum space.
We define a new set of reciprocal basis vectors
\begin{equation}
\begin{aligned}
B_1 = &  b_1= \frac{\sqrt{3} \pi}{4}(1,0,0),\\
B_2 = & b_2+4b_1 = \sqrt{3}\pi (0,1,0),\\ 
B_3 = & b_3+b_2= \sqrt{3}\pi  (0,0,1)\\
\end{aligned}
\end{equation}

The BZ spanned by above basis vectors is cuboid, instead of the BZ with irregular shape in Eq.~\eqref{b_nu}.
This definition make it convenient for the summation in Eq.~\eqref{k_xy},
The BZ is also used in the plot of band structure in Fig.~\ref{3} and Berry curvatures in Fig.~\ref{4}, 
and the calculation of Chern number in Eq.~\eqref{Berry_Chn}.

\bibliography{Ref}

\end{document}